\documentclass[aps,pra,reprint,superscriptaddress,showpacs]{revtex4-1}

\usepackage[]{hyperref} 
\usepackage[cp1250]{inputenc}
\usepackage{amsmath}
\usepackage{amsfonts}
\usepackage{textcomp}
\usepackage[]{graphicx}
\usepackage{mathrsfs}
\usepackage{amsbsy}
\usepackage{cleveref}
\usepackage[rightcaption]{sidecap}

\DeclareMathOperator{\sech}{sech}

\hypersetup{pdfauthor={Wiktor Walasik, Gilles Renversez, Yaroslav Kartashov},pdftitle={Stationary plasmon--soliton waves in metal--dielectric nonlinear planar structures: modeling and properties}}

\begin{document}

\title{Stationary plasmon--soliton waves in metal--dielectric nonlinear planar structures: modeling and properties}

\author{Wiktor Walasik}
\affiliation{Aix--Marseille Universit\'{e}, CNRS, Centrale Marseille, Institut Fresnel, UMR 7249, 13013 Marseille, France}
\email[]{gilles.renversez@fresnel.fr, www.fresnel.fr/spip/clarte}
\affiliation{ICFO --- Institut de Ci\`{e}ncies Fot\`{o}niques, Universitat Polit\`{e}cnica de
Catalunya, 08860 Castelldefels (Barcelona), Spain}
\author{Gilles Renversez}
\affiliation{Aix--Marseille Universit\'{e}, CNRS, Centrale Marseille, Institut Fresnel, UMR 7249, 13013 Marseille, France}
\author{Yaroslav Kartashov}
\affiliation{Institute of Spectroscopy, Russian Academy of Sciences, Troitsk, Moscow Region 142190, Russia}

\date{\today}

\begin{abstract}
We present three complementary methods to study stationary nonlinear solutions in one-dimensional nonlinear metal--dielectric structures. Two of them use an approximate treatment of the Kerr type nonlinear term taking into account only the leading electric field component, while the third one allows for an exact treatment of the nonlinearity.
A direct comparison
of the results obtained with all three models is presented and the excellent agreement between them
 justifies the assumptions that have been used to
construct the models.  
 A systematic study of the configurations made of two, three, or four layers, that
contain a semi-infinite Kerr type nonlinear dielectric, a metal film and linear dielectrics is presented. 
Detailed analysis of properties, type and
number of solutions in these three types of structures is performed. The parameter ranges where plasmon--soliton waves exist are found.
The structures with realistic opto-geometric parameters where plasmon--solitons exist at power levels already used in spatial soliton studies are proposed and studied.
\end{abstract}

\pacs{42.65.Wi, 42.65.Tg, 42.65.Hw, 73.20.Mf}
\keywords{Nonlinear waveguides, optical, Optical solitons, Kerr effect: nonlinear optics, Plasmons on surfaces and interfaces / surface plasmons}

\maketitle

\section{Introduction}

Stationary nonlinear waves coupling together a surface plasmon and a spatial optical soliton have been under investigation since at least 1985, when they were constructed for the first time by Ariyasu \textit{et~al.}~\cite{Ariyasu85} using a semi-analytical approach first suggested by Agranovich \textit{et~al.}~in 1980~\cite{Agranovich80} and used also in Ref.~\cite{Stegeman84}. Many theoretical and numerical works on this type of nonlinear waves followed~\cite{Tomlinson80, Maradudin81, Stegeman84c, Akhmediev82, Stegeman85, Seaton85, Boardman86, Prage91,  Stegeman84II, Mihalache85, Boardman87, Mihalache87, Chen88}. For reviews one can refer to Refs~\cite{Maraduduin83, Mihalache89}. In 2007, Feigenbaum and Orenstein~\cite{Feigenbaum07} coined the term 'plasmon--soliton' to name the wave propagating in a metal slot waveguide with a Kerr type nonlinear dielectric core. More recently, linked to the growth of the plasmonics research field~\cite{Maier07,Rukhlenko11}, several articles studied plasmon--solitons in details using various approaches~\cite{Bliokh09,Yin09,Davoyan09,Marini11,Milian12,Walasik12,Kou12,Ferrando13}. Nevertheless, in spite of all these results, no experimental observation of plasmon--solitons has been published yet, due to the too high power or, equivalently, too high nonlinear refractive index change required to generate the coupling between the plasmon and the soliton. This issue has been solved, at least theoretically but using realistic parameters, in our recent letter~\cite{Walasik12}. The proposed planar structure is made of a bulk nonlinear dielectric substrate covered by a thin linear dielectric film with a refractive index lower than that of the nonlinear medium and  a thin metal layer on top, in contact with a low-index linear external medium. It was shown that such structure supports plasmon--solitons with a peak power density as low as 1~GW/cm$^2$. This level of power density was already used experimentally to generate and record spatial solitons in fully dielectric planar chalcogenide waveguides~\cite{Chauvet09}.

In the present work, the details of the method used in Ref.~\cite{Walasik12} (which was based in the approach proposed in Refs.~\cite{Ariyasu85,Agranovich80}) are provided and two complementary approaches that confirm the validity of these results are also described. The first one is a semi-analytical approach that does not require any hypothesis concerning the field shape except for zero value boundary conditions at infinity and that takes into account both longitudinal and transverse components of the electric field in the Kerr type nonlinear term. This method extends to four-layer configuration the work published by Yin \textit{et~al.}~\cite{Yin09}. The second approach, more numerical, is based on a finite element method that finds iteratively both the field profile and the propagation constant of the nonlinear stationary waves. 

Concerning the results, this work is limited to one-dimensional stationary nonlinear solutions since they represent a necessary preliminary step to more complicated studies like temporal evolution and stability analysis or two-dimensional problems. For a review of all-dielectric nonlinear planar structures the reader can refer to Refs.~\cite{Akhmediev97,Kivshar03}. Our work is also limited to the focusing Kerr type nonlinearity as most of the materials used in integrated optics with third order nonlinearity are of this type~\cite{Martelucci94}. New and complete results for planar structures containing two, three or four layers are given including the type and the number of nonlinear solutions as a function of the opto-geometric parameters of the structure.
These results illustrate that low-power plasmon--solitons were not found previously mainly due to the limited parameter region in which they exist. They also prove that the simplest planar structures supporting low-power plasmon--solitons with a pronounced soliton peak and with a plasmonic part in a low-index external medium like air or water should involve four layers, as suggested in Ref.~\cite{Walasik12}.

The outline of the article is the following. After the statement of the problem in section~\ref{sec:problem}, three different approaches to compute its solutions are described in section~\ref{sec:models}. These methods are validated through comparisons with already published results and through mutual comparisons in sections~\ref{sec:ver} and \ref{sec:results}. In section~\ref{sec:results} the properties of solutions in two-, three-, and four-layer configurations are described with a more detailed analysis for the last one.

\section{Problem statement}
\label{sec:problem}

\begin{figure}[!ht]
\centering
\includegraphics[width = 0.48\textwidth,angle=-0,clip=true,trim=0 0 0 0]{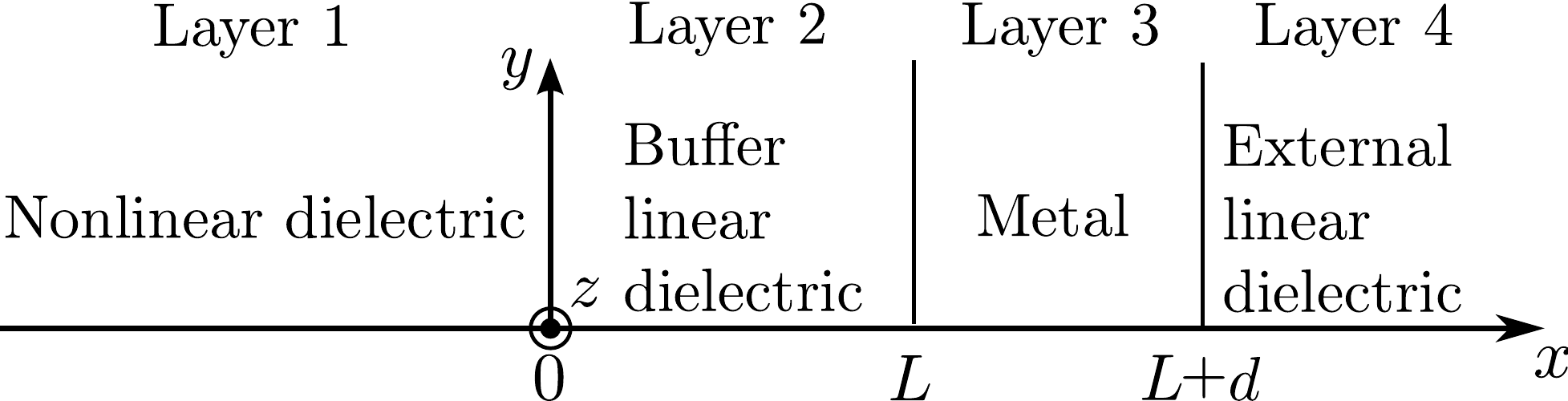}
\caption{Geometry of the one-dimensional four-layer nonlinear configuration.}
\label{fig:geometry}
\end{figure}

In this article we present three methods based on Maxwell's equations, to study the properties of stationary solutions in one-dimensional structures composed of semi-infinite nonlinear medium and layers of metal and linear dielectrics as depicted in Fig.~\ref{fig:geometry}. In all the described approaches only transverse magnetic (TM) polarized waves are considered due to the presence of the metal layer. We found no localized solutions for the transverse electric (TE) polarisation.

The first model extends and modifies the approach presented in Ref.~\cite{Ariyasu85} and uses two assumptions: (i) the nonlinearity depends only on the transverse electric field component and (ii) the nonlinear permittivity modifications are low compared to the linear part of the permittivity. These assumptions allow to write a single nonlinear wave equation for one of the magnetic field components. This equation is then solved analytically~\cite{Agranovich80}, resulting in closed formulas for the dispersion relation and for the electro-magnetic field shapes. This model will be called 'field based model' (FBM).

The second model, named in this article the 'exact model' (EM) because it does not require any of the two above assumptions, is based on the approaches from Refs.~\cite{Mihalache87,Yin09}. It also provides a closed formula for the nonlinear dispersion relation but the field shapes in the nonlinear medium are not given in an analytical form and have to be computed numerically by solving a system of two coupled first order nonlinear differential equations. 

The third model --- in contrast with the two previous semi-analytical ones --- uses a numerical finite element method (FEM) to solve the nonlinear scalar TM problem in layered structures. This approach finds the solutions using the fixed power algorithm from Refs.~\cite{Ettinger91,Li92, Drouart08} adapted to one-dimensional planar metal--dielectric structures.  

Our models are written for TM light polarization where the magnetic field has only one component $\pmb{\mathscr{H}} = [0,\mathscr{H}_y,0]$ and the electric field has two components $\pmb{\mathscr{E}} = [\mathscr{E}_x,0,i \mathscr{E}_z]$. The stationary solutions in one-dimensional geometry are sought in the form of monochromatic harmonic waves:
\begin{equation}
\begin{Bmatrix}
\pmb{\mathscr{E}}(x,z,t)\\ 
\pmb{\mathscr{H}}(x,z,t)
\end{Bmatrix} = 
\begin{Bmatrix}
\textbf{E}(x)\\ 
\textbf{H}(x)
\end{Bmatrix} e^{i( k_0 \beta z - \omega t)}.
\label{eqn:harmonic1}
\end{equation}
The propagation direction is chosen to be $z$ and $\omega$ denotes the angular frequency of the wave. $k_0 = \omega / c$ denotes the wavenumber in vacuum, $c$ denotes the speed of light in vacuum, and $\beta$ denotes the effective index of the propagating wave (the propagation constant is expressed as $k_0 \beta$). The structure is invariant along the $y$ direction and therefore it is assumed that the field shapes are invariant along the $y$ coordinate.

In this work a nonlinear Kerr type dielectric is considered in which the permittivity depends on the electric field intensity $\epsilon = \epsilon_l + \alpha |\textbf{E}|^2.$ Only the case of a focusing nonlinearity ($\alpha > 0$) is studied.
The relation between the nonlinear parameter $\alpha$ and the coefficient $n_2$, that appear in the definition of an intensity dependent refractive index $n = n_0 + n_2 I$, is $\alpha  = \epsilon_0 c \epsilon_{l} n_2$  (for $n_2 I \ll n_0$), where the intensity is defined as $I = \epsilon_0 c \sqrt{\epsilon_l} |\textbf{E}|^2 /2$ \cite{Boyd07}, $n_0$ denotes the linear part of the refractive index, and the vacuum permittivity is denoted by $\epsilon_0$.

\section{Model derivation}
\label{sec:models}

\subsection{Maxwell's equations}

The derivation of our models starts from the general form of the Maxwell's equations in case of nonmagnetic materials (relative permeability~$\mu = 1$) without free charges ($\rho_f = 0$) and free currents ($\textbf{J}_f = 0$)~\cite{Jackson99}:
\begin{subequations}
\label{eqn:maxwell1}
\begin{align}
\nabla \times \pmb{\mathscr{E}}  &= -\frac{\partial \pmb{\mathscr{B}}}{\partial t},     \label{eqn:rotE1}\\
\nabla \times \pmb{\mathscr{H}}  &= \frac{\partial \pmb{\mathscr{D}}}{\partial t},     \label{eqn:rotH1} \\
\nabla \cdot \pmb{\mathscr{D}}  &= 0     \label{eqn:divD1}, \\
\nabla \cdot \pmb{\mathscr{B}}  &= 0     \label{eqn:divB1}.
\end{align}
\end{subequations}
The magnetic induction vector is defined as $\pmb{\mathscr{B}} = \mu_0 \pmb{\mathscr{H}}$ and the displacement vector is $\pmb{\mathscr{D}}~=~\epsilon_0 \tilde{\overline{\overline{\epsilon}}} \pmb{\mathscr{E}}$. The vacuum permeability is denoted by~$\mu_0$ and the relative complex permittivity tensor is assumed to be diagonal with isotropic losses:
\begin{equation}
\tilde{\overline{\overline{\epsilon}}} = \overline{\overline{\epsilon}} + i\; \overline{\overline{\epsilon}}'' =  \begin{pmatrix}
 \epsilon_x & 0 & 0\\ 
0 & \epsilon_y & 0\\ 
0 & 0 & \epsilon_z
\end{pmatrix} + i \begin{pmatrix}
 \epsilon'' & 0 & 0\\ 
0 & \epsilon'' & 0\\ 
0 & 0 & \epsilon''
\end{pmatrix}, 
\label{eqn:tensor}
\end{equation}
where $\epsilon_j$ ($j \in \{x,y,z\}$) and $\epsilon''$ are real quantities.

To derive the nonlinear dispersion relations for our structure we use only the real part of the permittivity tensor $\overline{\overline{\epsilon}}$ as in e.g., Refs.~\cite{Ariyasu85,Agranovich80,Tomlinson80, Maradudin81, Stegeman84c,Yin09} (the imaginary part will be used later in the calculation of losses). Equations~(\ref{eqn:rotE1}) and (\ref{eqn:rotH1}) are written for TM light polarization. Using the definitions of the magnetic induction and the displacement vector, Eq.~(\ref{eqn:rotE1}) gives
\begin{subequations}
\begin{equation}
i\frac{\partial \mathscr{E}_z}{\partial x} - \frac{\partial \mathscr{E}_x}{\partial z} = \mu_0 \frac{\partial \mathscr{H}_y}{\partial t}
\label{eqn:rotE3}
\end{equation} 
and Eq.~(\ref{eqn:rotH1}) yields
\begin{align}
 - \frac{\partial \mathscr{H}_y}{\partial z} &= \epsilon_0\epsilon_x \frac{\partial \mathscr{E}_x}{\partial t},
\label{eqn:rotH31}\\
 \frac{\partial \mathscr{H}_y}{\partial x} &= i\epsilon_0\epsilon_z \frac{\partial \mathscr{E}_z}{\partial t},
\label{eqn:rotH32}
\end{align} 
\end{subequations}
where the $x$, $z$, and time dependencies are skipped in the field components and the permittivity to simplify the notation.
Using Eq.~(\ref{eqn:harmonic1}) the $z$ and time derivatives are eliminated from Eqs.~(\ref{eqn:rotE3})--(\ref{eqn:rotH32}) to finally give
\begin{subequations}
\label{eqn:max4}
\begin{align}
k_0 \beta E_x - \frac{dE_z}{d x} &= \omega \mu_0 H_y,
\label{eqn:rotE4}\\
E_x &= \frac{\beta}{\epsilon_0 \epsilon_x c} H_y,
\label{eqn:rotH41}\\
E_z &= \frac{1}{\epsilon_0 \epsilon_z \omega} \frac{d H_y}{d x}.
\label{eqn:rotH42}
\end{align} 
\end{subequations}

The nonlinearity is of the isotropic Kerr type so that all the elements of the permittivity tensor depend in the same way on the electric field intensity in the nonlinear medium
\begin{equation}
\epsilon_j(x) = \epsilon_{l,j}(x) + \epsilon_{\textrm{nl}}(x),
\label{eqn:perm_nonl}
\end{equation} 
where $j \in \{ x,y,z \}$ and $\epsilon_{l,j}$ denotes the linear, real part of the permittivity, $\epsilon_{\textrm{nl}} = \alpha(x)|\textbf{E}(x)|^2$ denotes the nonlinear part of the permittivity limited to the optical Kerr effect that depends on the electric field intensity, and $\alpha(x)$ denotes the function that takes values of the nonlinear parameters associated with different layers (in linear materials it is null).

\subsection{Field based model}
\label{sec:field_model}
\subsubsection{Nonlinear wave equation}
\label{sec:nl-wv-eq}

From Eqs.~(\ref{eqn:max4}), the problem of finding stationary solutions using a nonlinear wave equation is formulated. The derivation presented here is similar to the one proposed by Agranovich \textit{et~al.}~\cite{Agranovich80}, in which the first description of the nonlinear localized surface plasmon polariton waves was given. In this seminal paper, the analytical expressions for the dispersion relation and for the field shapes of the nonlinear solutions at a single metal/nonlinear dielectric interface were found in a TM case using the assumption that only two of the permittivity tensor elements depend on the longitudinal electric field component~\footnote{In the geometry from Ref.~\cite{Agranovich80} $x$ direction is the longitudinal one contrarily to our notations where $z$ is the longitudinal direction.} $\epsilon_x = \epsilon_y = \epsilon_l  +\alpha |E_x|^2$, where nonlinearity is defocussing ($\alpha<0$). Later on, this model was improved by introducing more realistic assumptions on the nonlinear term (e.g.\;focusing nonlinearity depending only on the transverse component of the electric field~\cite{Stegeman85, Maraduduin83}). It was also extended to consider TE polarized waves as well as focusing and defocussing Kerr nonlinearities~\cite{Maradudin81, Stegeman84c}. Furthermore, the model of Agranovich \textit{et~al.} was expanded to consider nonlinear waves guided by a thin metal film sandwiched between nonlinear dielectrics~\cite{Stegeman84, Ariyasu85, Seaton85, Boardman86,  Prage91}. 

Our FBM improves and extends previous approaches in three ways: (i) it improves the nonlinearity treatment so that all the diagonal elements of the permittivity tensor depend on the electric field in a nonlinear manner [Eq.~(\ref{eqn:perm_nonl})], (ii) it improves the way the nonlinearity is taken into account in the dispersion relation derivation [Eqs.~(\ref{eq:disp_full})--(\ref{eqn:q-nl})] and in the electric field shapes calculations (section~\ref{sec:field-exp}), and (iii) it extends the existing model from three-layer structure to four-layer structure (the benefits of using four-layer structures are discussed in section \ref{sec:4-lay}). The derivation given below follows the lines presented in Ref.~\cite{Ariyasu85} with the improvements mentioned above.

Taking the derivative of Eq.~(\ref{eqn:rotH42}) with respect to $x$ and using Eqs.~(\ref{eqn:rotE4}), (\ref{eqn:rotH41}), and (\ref{eqn:perm_nonl}) gives
\begin{equation}
\frac{d^2 H_y}{d x^2} = k_0^2 \left(\frac{\epsilon_z}{\epsilon_x} \beta^2 - \epsilon_z\right) H_y +\epsilon_0 \omega \frac{d \epsilon_{\textrm{nl}}}{d x} E_z.
\label{eqn:wave1}
\end{equation} 
Making use of Eq.~(\ref{eqn:rotH42}) in the last term allows to eliminate both electric field components from the equation, yielding an equation for the magnetic field component:
\begin{equation}
\frac{d^2 H_y}{d x^2} = k_0^2 \left(\frac{\epsilon_z}{\epsilon_x} \beta^2 - \epsilon_z\right) H_y + \frac{1}{\epsilon_z} \frac{d \epsilon_{\textrm{nl}}}{d x} \frac{d H_y}{d x}.
\label{eqn:wave2}
\end{equation} 

At this point an important assumption about the FBM is made. It is assumed that the nonlinear contribution to the permittivity is small compared to the linear part of permittivity $\epsilon_{\textrm{nl}} \ll \epsilon_{l,j}$ for $j \in \{x,z\}$ and both $\epsilon_{\textrm{nl}}$ and $H_y$ in the nonlinear medium vary in $x$ direction on scales larger than the wavelength. These hypothesis are valid for low-power solutions and are verified \textit{a posteriori} by analysing the field profiles. If they are fulfilled the last term in Eq.~(\ref{eqn:wave2}) is small and it can be omitted. Then the nonlinear wave equation can be written in the form:
 \begin{equation}
\frac{d^2 H_y}{d x^2} = k_0^2 \left(\frac{\epsilon_z}{\epsilon_x} \beta^2 - \epsilon_z\right) H_y .
\label{eqn:wave3}
\end{equation} 
The approximation made above affects only solutions in the nonlinear layer. Solutions in the linear layers are calculated in an exact way. 

In the following, only materials with equal linear parts of the permittivity tensor elements $\epsilon_{l,x} =\epsilon_{l,z} \equiv \epsilon_l$ are studied. The nonlinearity considered for our FBM is of the usual Kerr type, where only the transverse electric field component $E_x$ contributes to the nonlinear response (this component is usually much stronger than the longitudinal component in the studied photonic structures~\cite{Stegeman85})
\begin{equation}
\epsilon_x(x) = \epsilon_z(x) = \epsilon(x) =  \epsilon_l (x)+ \alpha(x) E_x^2(x).
\label{eqn:kerr_simple}
\end{equation} 
Using this form of nonlinearity, Eq.~(\ref{eqn:rotH41}), and again the assumption that  $\epsilon_{\textrm{nl}} \ll \epsilon_{l}$ [this assumption justifies the substitution of ${k_0 \beta H_y}/(\epsilon_0 \epsilon_x \omega) $ by ${k_0 \beta H_y}/(\epsilon_0 \epsilon_l \omega)$ in the nonlinear term] the nonlinear wave equation can be rewritten in its final form:
\begin{equation}
\frac{d^2 H_y}{d x^2} - k_0^2 q(x)^2 H_y + k_0^2 a(x) H_y^3 = 0,
\label{eqn:wave4}
\end{equation} 
where 
\begin{equation}
q(x)^2 = \beta^2 - \epsilon_l(x)
\label{eqn:q-def}
\end{equation}
and $a(x) =\beta^2 \alpha(x) /[\epsilon_0 \epsilon_l(x) c]^2$ is nonzero only in the nonlinear layer. $\epsilon_l(x)$ and $\alpha(x)$ are step-wise functions which take values indicated in Table \ref{tab:eps} depending on the layer as presented in Fig.~\ref{fig:geometry}.
\begin{table}[ht]
\centering
  \begin{tabular}{| c | c | c | c | c |}
    \hline
    Layer & Abscissa & $\epsilon_l(x)$  & $\epsilon(x)''$ & $\alpha(x)$ \\ \hline
    1 & $x<0$             &  $\epsilon_{l,1}$  & $\epsilon_1''$      & $\epsilon_0 c \epsilon_{l,1} n_2^{(1)} = \alpha_1$ \\ 
    2 & $0\le x< L $     &  $\epsilon_2 = \epsilon_{l,2}$        & $\epsilon_2''$      & 0 \\
    3 & $L\le x< L+d $ &  $\epsilon_3 = \epsilon_{l,3}$        & $\epsilon_3''$      & 0 \\
    4 & $ x\ge L+d $    &  $\epsilon_4 = \epsilon_{l,4}$        & $\epsilon_4''$      & 0 \\
    \hline
  \end{tabular}
\caption{Values of the functions describing the properties of the materials in different layers. The second-order nonlinear refractive index in layer 1 is denoted by $n_2^{(1)}$.}
  \label{tab:eps}
\end{table}

Equation (\ref{eqn:wave4}) is equivalent to Eqs.~(4) in Ref.~\cite{Ariyasu85} and to Eq. (14) in Ref.~\cite{Stegeman85} with a slight difference in the nonlinear term due the to more consistent nonlinearity treatment used here. The nonlinear function $a(x)$ differs by a factor $\beta^2/ \epsilon_l$ between our approach and the approaches from Refs.~\cite{Stegeman85,Ariyasu85}. This  results in discrepancies between our model and the older models mainly when the effective index of the nonlinear wave is much higher than the linear part of the nonlinear medium refractive index.

\subsubsection{Dispersion relation}
\label{sec:disp_rel_FBM}

The FBM provides solutions of the nonlinear wave equation [Eq.~(\ref{eqn:wave4})] for the $H_y$ field component. The solutions of this equation are studied separately in each layer of the structure. Then the use of boundary and continuity conditions allows us to obtain the nonlinear dispersion relation for the studied problem.

The solution of  Eq.~(\ref{eqn:wave4}) is well known in the literature~\cite{Agranovich80, Maradudin81,Lederer_83,Lederer_83II}.
In the nonlinear layer the solution is in the form (the $y$ subscript of the magnetic field is skipped as in our models there is only one magnetic field component, while the subscript~1 indicates the nonlinear layer, see Fig.~\ref{fig:geometry}):
\begin{subequations}
\label{eqn:H_open}
\begin{align}
 H_1 &= \sqrt{\frac{2}{a_1}} \frac{q_1}{\cosh[k_0 q_1 (x - x_0)]}   \;\;\;\; \;\;\;\;& \textrm{for } x<0, \label{eqn:H_nl}
\end{align}
where the $x_0$ is a free integration parameter that can be arbitrarily chosen and $q_k$ and $a_k$ denote the constant value of the $q(x)$ and $a(x)$ functions in the $k$-th layer.
If $x_0$ is negative, is has a physical meaning of the soliton peak position in the nonlinear dielectric. If it is positive, there is no maximum of the $H_1$ component in this layer.

In linear layers the nonlinear term in Eq.~(\ref{eqn:wave4}) vanishes and the solutions of the linear wave equation are expressed in a standard form of decreasing and increasing exponentials (for the layer indices see Fig.~\ref{fig:geometry}):
\begin{align}
 H_2 &= A_+ e^{k_0 q_2 x} +  A_- e^{-k_0 q_2 x}     \;\;\;\;    \;\;\;\;\textrm{for }  0 \le x < L, \\
 H_3 &= B_+ e^{k_0 q_3 (x - L)}  \nonumber \\ &+ B_- e^{-k_0 q_3 (x - L)}  \;\;\;\;\; \;\;\;\; \;\;\;\;   \textrm{for }   L \le x < L+d, \\ 
 H_4 &= C e^{-k_0 q_4 [x-(L+d)]}  \;\;\;\;  \;\;\;\;  \;\;\;\;  \;\;\;\;  \textrm{for } x \ge L+d.     \label{eqn:H_4}
\end{align}
\end{subequations} 
The use of the boundary condition $H_y \xrightarrow[]{x\rightarrow \infty}\nolinebreak0$  in the layer 4 results in the single term in Eq.~(\ref{eqn:H_4}).

Finally, using the conditions for the continuity of the $H_y$ and $E_z$ fields at the interfaces [$E_z$ is calculated using Eq.~(\ref{eqn:rotH42})], the analytical form of the nonlinear dispersion relation of the four-layer model is obtained:
\begin{subequations}
\label{eq:disp_full}
\begin{equation}
\Phi_{+}  \Big(\widetilde{q_4} + \widetilde{q_3} \Big) \exp(2k_0 \widetilde{q_3} \epsilon_3 d) + \Phi_{-} \Big(\widetilde{q_4} - \widetilde{q_3} \Big) =0, \;\; 
\label{eq:disp}
\end{equation}
where
\begin{equation}
\Phi_{\pm} = \bigg(  1  \pm\frac{\widetilde{q_{1,\textrm{nl}}}|_{x=0}}{\widetilde{q_3}}   \bigg)  + \bigg( \frac{\widetilde{q_{1,\textrm{nl}}}|_{x=0}}{\widetilde{q_2}}  \pm \frac{\widetilde{q_{2}}}{\widetilde{q_3}} \bigg) \tanh(k_0 q_{2} L) ,
\label{eqn:phi}
\end{equation}
\end{subequations}
and
\begin{align}
\widetilde{q_k}& = \frac{q_k}{\epsilon_k} \hspace{5.5em} \textrm{for}\hspace{0.3em}  k \in \{2,3,4\},\\
\label{eqn:q1nltilde}
\widetilde{q_{1,\textrm{nl}}} &=  \widetilde{q_1} \tanh(k_0 q_1 x_0).
\end{align}
Some assumptions have to be made in order to obtain the closed form of the expression for $\widetilde{q_1}$ and therefore of the nonlinear dispersion relation.
The exact expression for $\widetilde{q_1}$ reads
\begin{equation}
\widetilde{q_1} = \frac{q_1}{\epsilon_1} = \frac{q_1}{\epsilon_{l,1} + \alpha_1 E_x^2}
\end{equation} 
Here the model presented by Ariyasu \textit{et~al.} is improved once again. In Ref.~\cite{Ariyasu85} the nonlinear term is omitted at this step and $\widetilde{q_1} =  {q_1}/{\epsilon_{l,1}}$. Nevertheless, one can go beyond and find the first order approximation for $\widetilde{q_1}$ taking into account the nonlinearity.
$\widetilde{q_1}$ is expressed in terms of the magnetic field $H_1$. Using Eq.~(\ref{eqn:rotH41})
\begin{equation}
\widetilde{q_1} =  \frac{q_1}{\epsilon_{l,1} + \alpha_1 \left(\frac{k_0 \beta}{\omega \epsilon_0 \epsilon_{l,1}}\right)^2 H_1^2},
\label{eqn:q1tilde}
\end{equation}
where at this stage the assumption that $\epsilon_1 = \epsilon_{l,1}$ was used in the nonlinear term in the denominator of Eq.~(\ref{eqn:q1tilde}). Use of Eq.~(\ref{eqn:H_nl}) and the definition of $a(x)$ function results in
\begin{equation}
\widetilde{q_1} =  \frac{q_1}{\epsilon_{l,1} + 2 q_1^2\sech^2[k_0 q_1 (x-x_0)]}.
\end{equation}

To obtain the dispersion relation we need to know the value of $\widetilde{q_{1,\textrm{nl}}}$ at the interface $x=0$ which is
\begin{equation}
\widetilde{q_{1,\textrm{nl}}}|_{x=0} =  \frac{q_1  \tanh(k_0 q_1 x_0)}{\epsilon_{l,1} + 2 q_1^2\sech^2 (k_0 q_1 x_0)}.
\label{eqn:q-nl}
\end{equation}
Now the dispersion relation (\ref{eq:disp_full}) depends only on the wavenumber ($k_0$), material and structure parameters ($\epsilon_{l,1}$, $\epsilon_2$, $\epsilon_3$, $\epsilon_4$, $L$, $d$), the $x_0$ parameter, and the effective index ($\beta$). By fixing the values of the material and geometric parameters and $x_0$, one obtains a nonlinear expression that is satisfied only for a limited set of $\beta$ values. We are interested only in the solutions with $\beta > \sqrt{\epsilon_{l,1}}$ because the solutions we look for should be localized either in the nonlinear dielectric or at the metal/nonlinear dielectric interface \{see the definition of $q$ [Eq.~(\ref{eqn:q-def})] and the field profiles [Eqs.~(\ref{eqn:H_open})]\}. It is worth noting that the dispersion relation does not depend on the nonlinear parameter $\alpha_1$. This is a consequence of the fact that the nonlinear solution depends on the nonlinear permittivity modification $\epsilon_{\textrm{nl}} \propto \alpha_1 E_x^2$ and not on the field amplitude or nonlinear parameter itself. Changing the nonlinearity coefficient does not result in change of the effective indices that verify the dispersion relation but only in change of the field amplitude as it can be seen by rescaling all the fields by a factor $\sqrt{\alpha_1}$.
 
\subsubsection{Power and losses}

In the FBM, having determined the effective indices $\beta$, a closed analytical expression for the approximated power density of the corresponding plasmon--soliton waves can be found. Power density transmitted per unit length along $y$ direction is expressed as a longitudinal ($z$) component of the pointing vector $\textbf{S} = \frac{1}{2}$Re$(\textbf{E} \times \textbf{H}^*)$ integrated over the transverse dimension ($x$)
\begin{equation}
P = \int_{-\infty}^{+\infty} S_z dx = \frac{1}{2} \int_{-\infty}^{+\infty} E_x H_y^* dx,
\end{equation}
which is rewritten using Eq.~(\ref{eqn:rotH41}) in the form:
\begin{equation}
P =  \frac{\beta}{2 c \epsilon_0} \int_{-\infty}^{+\infty} \frac{1}{\epsilon_x(x)} |H_y|^2 dx.
\label{eqn:Power_exact}
\end{equation}
In this expression the dependency of the permittivity on the $x$ coordinate is both due to a layered structure (linear) and due to the field induced changes in the nonlinear layer. If we use again the assumption that $\epsilon_{\textrm{nl}} \ll \epsilon_l$ then the expression for the total approximated power density $P$ can be rewritten as a sum of four separate integrals
\begin{equation}
P =  \sum_{k = 1}^4 P_k,
\end{equation}
where 
\begin{equation}
P_k =  \frac{\beta}{2 c \epsilon_0 \epsilon_{l,k}} \int_{\textrm{layer } k}  |H_k|^2 dx.
\label{eqn:Power_approx}
\end{equation}
Using Eq.~(\ref{eqn:H_open}) we find the approximate power density in the nonlinear layer
\begin{subequations}
\label{eqn:Power}
\begin{align}
P_1 = \frac{\beta}{2 c \epsilon_0 \epsilon_{l,1} k_0} \frac{ H_0^2}{q_1-q_{1,\textrm{nl}}},
\end{align}
where $q_{1,\textrm{nl}} = \widetilde{q_{1,\textrm{nl}}} \epsilon_{1}$, and the exact expressions for the power densities in the linear layers:
\begin{align}
P_2  &= \frac{\beta }{2 c \epsilon_0 \epsilon_2} \left [ \frac{A_+^2}{2 k_0 q_2} \left (e^{2 k_0 q_2 L} - 1 \right ) \right . \nonumber \\
         &   \left .  + 2A_+A_-L - \frac{A_-^2}{2 k_0 q_2} \left (e^{-2 k_0 q_2 L} - 1 \right ) \right ],\\
P_3 &= \frac{\beta }{2 c \epsilon_0 \epsilon_3} \left [ \frac{B_+^2}{2 k_0 q_3} \left (e^{2 k_0 q_3 d} - 1 \right ) \right . \nonumber \\
         & \left . + 2B_+B_-d - \frac{B_-^2}{2 k_0 q_3} \left (e^{-2 k_0 q_3 d} - 1 \right ) \right ],\\ 
P_4 &= \frac{\beta}{2 c \epsilon_0 \epsilon_4} \frac{C^2}{2 k_0 q_4},
\end{align}
\end{subequations} 
where $H_0,A_+,A_-,B_+,B_-$, and $C$ are found during the procedure of solving the nonlinear dispersion relation [Eqs.~(\ref{eq:disp_full})] and are given by:
\begin{subequations}
\label{eqn:coeff}
\begin{align}
H_0 &=  \sqrt{\frac{2}{a_1}}  \frac{q_1}{\cosh(k_0 q_1 x_0)}, \\
A_{\pm} &= \frac{H_0}{2} \left( 1\pm \frac{\widetilde{q_{1,\textrm{nl}}}|_{x=0}}{\widetilde{q_2}} \right),\\
B_\pm &= \frac{H_0}{2} \left[ \left( 1 \pm \frac{\widetilde{q_{1,\textrm{nl}}}|_{x=0}}{\widetilde{q_3}}  \right) \cosh(k_0 q_2 L)  \right . \nonumber \\
    &  \;\;\;\;\;\;\;\;\;\;   \left . + 
         \left(  \frac{\widetilde{q_{1,\textrm{nl}}}|_{x=0}}{\widetilde{q_2}} \pm \frac{\widetilde{q_2}}{\widetilde{q_3}}  \right) \sinh(k_0 q_2 L) \right], \\
C &= B_+ e^{k_0 q_3 d} + B_- e^{-k_0 q_3 d}.
\end{align}
\end{subequations}

An important part of the study of nonlinear wave propagation is the calculation of losses. In our FBM, the losses are estimated using the approach based on the imaginary part of permittivity and the field profiles [the complex permittivity function is denoted by $\tilde{\epsilon}(x) = \epsilon(x) + i \epsilon''(x)$ and it takes values given in Table~\ref{tab:eps}]. This method is described in the case of linear waveguides in Ref.~\cite{Snyder83} and has already been used for nonlinear plasmon--soliton studies~\cite{Maraduduin83,Stegeman85,Walasik12}.

The expression that provides an approximation of the imaginary part of the effective index $\beta''$ is~\cite{Snyder83}
\begin{equation}
\label{eqn:Beta_im1d_fin}
\beta'' = \frac{\epsilon_0 c}{4} \frac{ \int_{-\infty}^{+\infty} \epsilon''(x) |\textbf{E}|^2 dx}{P}.
\end{equation}
The imaginary part of the refractive index is connected with the losses in decibel per meter ($\mathfrak{L}$) in the following way~\cite{Zolla12}:
 \begin{equation}
\mathfrak{L} = \frac{40\pi}{\ln(10) \lambda} \beta'',
\end{equation}
where $\lambda$ it the free-space wavelength expressed in meters.

\subsubsection{Expressions for the electric field components}
\label{sec:field-exp}
In our FBM, the wave equation for the $H_y$ magnetic field component [Eq.~(\ref{eqn:wave4})] is solved and the analytical expressions for the field shape of this component are provided [Eqs.~(\ref{eqn:H_open})]. In the case of a linear medium, knowing the expression for magnetic field one can easily calculate the electric field components using  Eqs.~(\ref{eqn:rotH41}) and (\ref{eqn:rotH42}). In the nonlinear case this problem requires precautions. If the permittivity depends on the TM wave electric field components these two equations form a set of two coupled nonlinear equations
\begin{subequations}
\label{eqn:E_fields}
\begin{align}
E_x &= \frac{\beta}{\epsilon_0 \epsilon_x(E_x,E_z) c} H_y,
\label{eqn:Exset}\\
E_z &= \frac{1}{\epsilon_0 \epsilon_z(E_x,E_z) \omega} \frac{d H_y}{d x}.
\label{eqn:Ezset}
\end{align}
\end{subequations}
However, in the frame of the FBM, a simplified Kerr dependency for the permittivity is assumed through Eq.~(\ref{eqn:kerr_simple}), where the permittivity depends only on the main electric field component $E_x$. In this case the problem reduces to
\begin{subequations}
\label{eqn:E_fields2}
\begin{align}
E_x &= \frac{\beta}{\epsilon_0 \epsilon_x(E_x) c} H_y,
\label{eqn:Exset2}\\
E_z &= \frac{1}{\epsilon_0 \epsilon_z(E_x) \omega} \frac{d H_y}{d x}.
\label{eqn:Ezset2}
\end{align}
\end{subequations}
The first equation is no longer coupled to the second one and contains only one unknown quantity ($E_x$) so it can be readily solved. Inserting Eq.~(\ref{eqn:kerr_simple}) into Eq.~(\ref{eqn:Exset2}) and performing some simple algebra give
 \begin{equation}
\label{eqn:cubic}
E_x^3 + \frac{\epsilon_{l,1}}{\alpha_1} E_x - \frac{\beta H_y}{\epsilon_0 c \alpha_1} = 0
\end{equation}

This equation has in general three roots: one real and a pair of complex conjugate solutions. In our approach the $E_x$ field is assumed to be real so we choose the real root of this equation to be the field profile. The solution is in the form~\cite{Abramowitz72}:
 \begin{align}
E_x =& \left(\frac{w}{2} + \sqrt{\frac{v^3}{27} + \frac{w^2}{4}}\right)^{\frac{1}{3}} + \left(\frac{w}{2} - \sqrt{\frac{v^3}{27} + \frac{w^2}{4}}\right)^{\frac{1}{3}},
\end{align}
with $w = \beta H_y/(\epsilon_0 c \alpha_1)$ and $v = \epsilon_{l,1}/\alpha_1$. Having found the $E_x$ field shape, the $E_z$ dependency is calculated directly using Eq.~(\ref{eqn:Ezset2}).

In previous approaches~\cite{Ariyasu85,Stegeman85} the electric field was calculated using simplified formulas containing only the linear part of the refractive index
\begin{subequations}
\label{eqn:E_fields_lin}
\begin{align}
E_x &= \frac{\beta}{\epsilon_0 \epsilon_{l,x} c} H_y,
\label{eqn:Exsetlin}\\
E_z &= \frac{1}{\epsilon_0 \epsilon_{l,z} \omega} \frac{d H_y}{d x}.
\label{eqn:Ezsetlin}
\end{align}
\end{subequations}

\subsection{Exact model}
\label{sec:exact}
\subsubsection{First integral nonlinear medium treatment}

Below we present the derivation of the model that allows the exact treatment of the Kerr nonlinearity.
This derivation is based on the approaches presented firstly by Mihalache \textit{et~al.}~\cite{Mihalache87} for two-layer configurations, later extended to three-layer configurations and generalized to the case of power-law Kerr nonlinearity by Yin \textit{et~al.}~\cite{Yin09}. Here we limit ourselves to usual cubic nonlinearity, but we extend the approach to a four-layer configuration.

This derivation starts from the Maxwell's equations [Eqs.~(\ref{eqn:max4})]. In this approach the magnetic field is eliminated from these equations. The use of Eq.~(\ref{eqn:rotH41}) in Eqs.~(\ref{eqn:rotE4}) and (\ref{eqn:rotH42}) gives
\begin{subequations}
 \begin{align}
\frac{d E_z}{d x} &= \left( \beta k_0 - \frac{k_0}{\beta} \epsilon_x \right)E_x.
\label{eqn:yin1}\\
\frac{d (\epsilon_x E_x)}{d x} &=  \beta k_0  \epsilon_z E_z. 
\label{eqn:yin2}
\end{align} 
\end{subequations}
Equation (\ref{eqn:yin1}) is derived with respect to $x$ and the last term is replaced using Eq.~(\ref{eqn:yin2}) resulting in
 \begin{equation}
\frac{d^2 E_z}{d x^2} =  \beta k_0 \frac{d E_x}{d x}  - k_0^2 \epsilon_z E_z.
\label{eqn:yin3}
\end{equation} 
Multiplying Eq.~(\ref{eqn:yin3}) by $dE_z/dx$ and using Eq.~(\ref{eqn:yin1}) once more gives
 \begin{equation}
\frac{d^2 E_z}{d x^2}\frac{d E_z}{d x} =  \beta k_0 \frac{d E_x}{d x}\left( \beta k_0 - \frac{k_0}{\beta} \epsilon_x \right)E_x  - k_0^2 \epsilon_z E_z \frac{d E_z}{d x}.
\label{eqn:yin4}
\end{equation} 
In this approach a full Kerr dependency of the permittivity of the following form is assumed in the nonlinear layer [compare with Eqs.~(\ref{eqn:perm_nonl}) and (\ref{eqn:kerr_simple})]:
 \begin{equation}
\epsilon_x = \epsilon_z = \epsilon_1= \epsilon_{l,1} + \alpha_1(E_x^2 + E_z^2).
\label{eqn:kerr_full}
\end{equation} 
Inserting this definition into Eq.~(\ref{eqn:yin4}) one obtains
\begin{widetext}
 \begin{equation}
\frac{d^2 E_z}{d x^2}\frac{d E_z}{d x} =  (\beta k_0)^2 E_x \frac{d E_x}{d x} - k_0^2 \epsilon_{l,1} \left( E_x   \frac{d E_x}{d x}   +  E_z \frac{d E_z}{d x} \right) -  k_0^2 \alpha_1(E_x^2 + E_z^2) \left( E_x   \frac{d E_x}{d x}   +  E_z   \frac{d E_z}{d x} \right).
\label{eqn:yin5}
\end{equation} 
\end{widetext}
Integrating this equation by parts with respect to $x$ gives
 \begin{align}
\left( \frac{d E_z}{d x} \right)^2 =  (\beta k_0)^2 E_x^2 &-  k_0^2 \epsilon_{l,1} \left( E_x ^2  +  E_z ^2 \right) \nonumber \\ 
&-  k_0^2 \frac{\alpha_1}{2}(E_x^2 + E_z^2)^2 + C_0,
\label{eqn:yin6}
\end{align} 
where $C_0$ is the integration constant. $C_0$ is set to 0 taking into consideration the fact that a semi-infinite nonlinear medium is studied, where the electric fields $E_x$, $E_z$, and their derivatives vanish as $x\rightarrow -\infty$.
The final step of this derivation is to compare the RHS of Eq.~(\ref{eqn:yin6}) with the square of the RHS of Eq.~(\ref{eqn:yin1}). This comparison yields
 \begin{equation}
\left( \frac{\epsilon_1^2}{\beta^2} - 2\epsilon_1 \right) E_z^2  + \epsilon_{l,1} \left( E_x ^2  +  E_z ^2 \right) +   \frac{\alpha_1}{2}(E_x^2 + E_z^2)^2 =0,
\label{eqn:yin7}
\end{equation} 
which is the first step in order to obtain the nonlinear dispersion relation in the EM.

\subsubsection{Dispersion relation and field shapes}
\label{sec:yin_disp_field}

In the previous paragraph a method that allows the treatment of the nonlinearity in an exact manner, without the approximations used in the FBM, was presented.
Besides, there is no difficulty in solving Maxwell's equations in the linear layers. In the following, the two electric field components in these layers are expressed as a combination of increasing and decreasing exponents. In the buffer linear dielectric ($0 \le x < L$ --- layer 2)
\begin{subequations}
\begin{align}
 E_{x,2} &= A_{x} e^{k_0 q_2 x} +  B_{x} e^{-k_0 q_2 x}   \label{eqn:Ex2},\\
 E_{z,2} &= A_{z} e^{k_0 q_2 x} +  B_{z} e^{-k_0 q_2 x}    \label{eqn:Ez2},
\end{align}
\end{subequations} 
in the metal ($ L \le x < L+d$ --- layer 3)
\begin{subequations}
\begin{align}
 E_{x,3} &= C_{x} e^{k_0 q_3 (x - L)} +  D_{x} e^{-k_0 q_3 (x - L)} \label{eqn:Ex3},\\
 E_{z,3} &= C_{z} e^{k_0 q_3 (x - L)} +  D_{z} e^{-k_0 q_3 (x - L)}  \label{eqn:Ez3},
\end{align}
\end{subequations} 
in the external linear dielectric ($ x \ge L+d$ --- layer 4)
\begin{subequations}
\begin{align}
 E_{x,4} &= F_{x} e^{-k_0 q_4 [x-(L+d)]}   \label{eqn:Ex4},\\
 E_{z,4} &= F_{z} e^{-k_0 q_4 [x-(L+d)]}   \label{eqn:Ez4}.
\end{align}
\end{subequations} 

Several relations between field amplitudes in the linear layers are needed. Using Eq.~(\ref{eqn:yin2}) separately in each of the linear and uniform layers, relations between the amplitudes of the $x$ and $z$ components of the fields are found. Inserting Eqs.~(\ref{eqn:Ex2}) and (\ref{eqn:Ez2}) into Eq.~(\ref{eqn:yin2}) one obtains the relation for the field amplitudes in layer~2:
\begin{equation}
\beta (A_z e^{k_0 q_2 x} + B_z e^{-k_0 q_2 x}) = q_2 (A_x e^{k_0 q_2 x} - B_x e^{-k_0 q_2 x}).
\end{equation}
Because this equation has to be fulfilled for each value of $x\in]0,L[$ we separately solve the equations for the terms proportional to $e^{k_0 q_2 x}$ and $e^{-k_0 q_2 x}$. As a result one obtains
\begin{equation}
 \left\{ \begin{aligned}
A_x &= \frac{\beta}{q_2} A_z \label{eqn:Ax},\\
B_x &= -\frac{\beta}{q_2} B_z.
\end{aligned}
 \right.
\end{equation}
Applying a similar procedure to the expressions of the fields in other linear layers leads to
\begin{equation}
 \left\{ \begin{aligned}
C_x &= \frac{\beta}{q_3} C_z,\\
D_x &= -\frac{\beta}{q_3} D_z,
\end{aligned}
 \right.
\end{equation}
\begin{align}
F_x &= -\frac{\beta}{q_4} F_z \label{eqn:Ex}.
\end{align}

Using Eqs.~(\ref{eqn:Ax})--(\ref{eqn:Ex}) and the continuity conditions for the fields $E_z$ and $H_y$ [computed using Eq. (\ref{eqn:rotH41})] at the boundaries between layers, the longitudinal component of the electric field at the nonlinear interface $ E_{z,0}\equiv E_z(x=0^-)$ is expressed as a function of the total electric field amplitude at this interface $E_0$:
\begin{subequations}
\begin{align}
E_{z,0}^2 &= \frac{\left({\epsilon_2 \beta}/{q_2} \right)^2 }
             {\epsilon_{1,0}^2 \left[{(1+\phi)}/{(1-\phi)} \right]^2 + \left({\epsilon_2 \beta}/{q_2} \right)^2 } E_0^2,
\label{eqn:Ez}\\
\phi &= \frac{\Psi^+_+ e^{- k_0 q_2 L - k_0 q_3 d} + \Psi^-_- e^{- k_0 q_2 L + k_0 q_3 d}}
                {\Psi^+_- e^{ k_0 q_2 L + k_0 q_3 d} + \Psi^-_+ e^{ k_0 q_2 L - k_0 q_3 d}},
\label{eqn:phiYin}\\
\Psi^{ \textrm{sgn}(m)}_{ \textrm{sgn}(p)} &= \frac{{\epsilon_2}/{q_2}+m {\epsilon_3}/{q_3}}
                                          {{\epsilon_3}/{q_3}+ p {\epsilon_4}/{q_4}}, \hspace{1em} \textrm{where } \{m,p\} = \{1,-1 \},
\label{eqn:psi}
\end{align}
\end{subequations}
and
\begin{equation}
E_0 = (E_{x,0}^2 + E_{z,0}^2)^{1/2},
\label{eqn:int-amp}
\end{equation}
and an additional subscript 0 denotes values of functions at ($x = 0^-$).

Using Eqs.~(\ref{eqn:Ez}) and (\ref{eqn:int-amp}) to eliminate $E_x$ and $E_z$ from Eq.~(\ref{eqn:yin7}) taken at $x = 0^-$ results in the final form of the nonlinear dispersion relation for the EM:
\begin{align}
\left(\frac{\epsilon_{1,0} \epsilon_2 }{q_2} \right)^2 &-
 2\epsilon_{1,0} \left(\frac{ \epsilon_2 \beta}{q_2} \right)^2 +\left( \epsilon_{l,1} + \frac{\alpha_1}{2} E_0^2 \right)  \nonumber \\
 &\times  \left[ \epsilon_{1,0}^2 \left( \frac{ 1 +\phi } {1 -\phi }\right)^2 \right .  \left . + 
 \left(\frac{\epsilon_2 \beta}{q_2} \right)^2    \right] = 0.
\label{eqn:Yin-disp}
\end{align}
For a given set of opto-geometric parameters and a given wavelength, it contains as a unique free parameter the total electric field amplitude at the nonlinear interface $E_0$. Fixing arbitrarily $E_0$ allows to solve this equation for all the possible values of $\beta$. 

After obtaining the effective indices of the nonlinear waves propagating in a given structure the field profiles corresponding to these values of $\beta$ must be calculated. In the EM, contrarily to the FBM, no analytical formulas for the field shapes in the nonlinear layer are provided. However, a system of two coupled first order differential equations for the electric field components can be derived to allow field shape computations. Eq.~(\ref{eqn:yin2}) is written in the form:
\begin{equation}
\frac{d \epsilon_x}{d x} E_x +\frac{d E_x}{d x} \epsilon_x  =  \beta k_0  \epsilon_z E_z. 
\end{equation} 
Using Eq.~(\ref{eqn:kerr_full}) in the first term and calculating the derivative give
\begin{equation}
2 \alpha_1 \left( E_x \frac{d E_x}{d x} + E_z \frac{d E_z}{d x} \right) E_x  +\frac{d E_x}{d x} \epsilon_1  =  \beta k_0  \epsilon_1 E_z. 
\end{equation} 
Replacing $dE_z/dx$ using Eq.~(\ref{eqn:yin1}) and reorganizing the terms result in the first coupled differential equation
\begin{align}
 \frac{d E_x}{d x} &= \frac{\beta k_0  \epsilon_1  E_z - 2  \alpha_1  E_z  E_x^2  \left(\beta k_0 - \frac{k_0}{\beta}  \epsilon_1 \right)}{ \epsilon_1 + 2  \alpha_1  E_x^2}.
\label{eqn:yin_fields_system2}
\end{align}
The second coupled differential equation used to calculate the field profiles is Eq.~(\ref{eqn:yin1}).

\subsection{Finite element method}

In this paragraph, the FEM based approach used to compute the stationary solutions propagating in the structure depicted in Fig.~\ref{fig:geometry} is described.
FEM has already been used to study stationary solutions in  nonlinear waveguides since at least the end of the eighties ~\cite{Hayata88, Rahman90, Rahman91}. 
For a general and recent review of finite element method in the frame of optical waveguides, the reader can refer to the  chapter 4 of Ref.~\cite{Zolla12}.
In the present case, the problem is relatively simple since it is both one-dimensional and is reduced to a scalar case.

The FEM is an approximation method of the solutions of differential partial equations. It is build from an equivalent formulation, the variational one, of the initial problem. 
 To get this new formulation also called weak formulation, the initial differential partial equations are multiplied by chosen form functions that belong to a particular function space  depending notably on the used  boundary conditions and the type of  differential partial equations. The next step to establish the FEM is the discretization, in which one shifts from an infinite dimension functional space to a finize size one that allow the numerical resolution.
It must be pointed out that the weak formulation  of the scalar problem for the full structure, deduced from Eq.~(\ref{eqn:wave3}) or its approximated form given by Eq.~(\ref{eqn:wave4}) must takes into account all the continuity relations fullfiled by the electromagnetic field at the structure interfaces.
This implies that the full TM wave equation for $H_y$ component  must be used  to obtain the correct weak formulation that deals with  both the inhomogeneous permittivity term induced by the nonlinearity  and the structure interfaces. 
The corresponding weak formulation is
\begin{widetext}
\begin{align}
-\int_{F}  \frac{1}{k^{2}_{0} \epsilon(x)}\mathbf{\nabla}\phi(x) \cdot\mathbf{\nabla}\phi'(x)dx&+\int_{F}\phi(x)\phi'(x)dx
=\beta^{2}\int_{F} \frac{1}{\epsilon(x)} \phi(x)\phi'(x)dx  \;\;\;\;\;\;\;
\forall\phi'\in\mathcal{H}^{1}_{0}(F) \; \mathrm{ and } \;  \phi\in\mathcal{H}^{1}_{0}(F)
\end{align}
\end{widetext}
 in which  $\mathcal{H}^{1}_{0}(F)$ is the Sobolev space of the order 1  with the null Dirichlet boundary conditions on the domain of integration $F$  (in the present case the full $x$ cross section of the structure). In the above equation   $\phi$ stands for the  $H_y$ component and $\phi'$ denotes the test form functions. The electric field components are calculated using Eqs.~(\ref{eqn:rotH41}) and (\ref{eqn:rotH42}) with the method described in section \ref{sec:field-exp}.
The FEM is implemented using the free softwares \textsc{Gmsh} as a mesh generator and \textsc{GetDP} as a solver~\cite{getdp-ieee1998,getdp-siam2008,Geuzaine09}. These softwares have already been used to solve both 2D scalar and vector nonlinear electromagnetic waveguide problems~\cite{Drouart08, Drouart-thesis09}. The nonlinearity considered in these two references was of the simplified Kerr type like in Eq.~(\ref{eqn:wave4}). 
The algorithm used for this plasmon--soliton study is the fixed power one~\cite{Ettinger91,Li92, Drouart08}  in which, for a given structure, the wave power is the input parameter and the outputs are  the propagation constant and the corresponding fields. This algorithm involves an iterative process requiring successive resolutions of generalized linear eigenvalue problems, where the square of the propagation constant $(k_0 \beta)^2$ is the eigenvalue and the field is the eigenvector. 
The iterative process is stopped when an arbitrary criterion on the convergence of the propagation constant is reached. Typically, $|(\beta_{n}-\beta_{n-1})/\beta_{n}| < \delta$, where $n$ denotes the step number in the procedure, and $\delta=10^{-6}$ is chosen in the present work. To fulfill this criterion between 10 and 15 steps are needed depending on the structure parameters and the used initial field. It is worth noticing that,  in the frame of the fixed power algorithm,
different initial fields provide at the end of the iterative process the same results except if the structure exhibits multiple solutions for the same power. In this last case, the obtained solution at the end of the  iterative process  depends on the initial field.

\section{Limiting cases for semi-analytical models}
\label{sec:ver}

\subsection {Field based model}
\label{sec:ver-fbm}
In order to verify our analytical results for the FBM, several comparisons with the formulas from previous works for simpler structures are realized in this section. 
The dispersion relations obtained in the frame of the FBM are considered in three limiting cases:
\subsubsection{Three-layer structure}

Assuming that $L~\rightarrow~0$, one notices immediately that $\tanh(k_0 q_{2} L)~\rightarrow~0$ and Eq.~(\ref{eqn:phi}) simplifies to
\begin{align}
\Phi_{\pm} = \bigg(  1\pm\frac{\widetilde{q_{1,\textrm{nl}}}|_{x=0}}{\widetilde{q_3}}   \bigg).
\end{align}
Inserting this expression into Eq.~(\ref{eq:disp}), after some simple algebra, yields
\begin{align}
\tanh(k_0 q_3 d) = - \frac{\widetilde{q_3}( \widetilde{q_{1,\textrm{nl}}}|_{x=0} + \widetilde{q_4})}
					 {\widetilde{q_3}^2 + \widetilde{q_4} \widetilde{q_{1,\textrm{nl}}}|_{x=0}}.
\label{eqn:disp_3nl}
\end{align}
If $ \widetilde{q_{1,\textrm{nl}}}|_{x=0}$ is approximated by $q_1\tanh(k_0 q_1 x_0)/\epsilon_{l,1}$ (see also section \ref{sec:disp_rel_FBM}) then the above equation is identical to Eq. (8) in Ref.~\cite{Ariyasu85} for the case of the three-layer structure, where the metal film is sandwiched between linear and nonlinear dielectrics.

\subsubsection{Two-layer structure}

An elegant way of finding the dispersion relation for two-layer structure is to infinitely separate both interfaces of the three-layer structure. This is done by letting $d~\rightarrow~\infty$. Then $\tanh(k_0 q_3 d) ~\rightarrow~1$ and Eq.~(\ref{eqn:disp_3nl}) becomes
\begin{align}
(\widetilde{q_{1,\textrm{nl}}}|_{x=0} + \widetilde{q_3}) (\widetilde{q_3} + \widetilde{q_4}) = 0. 
\end{align}
This equation has two solutions. The first one,
\begin{align}
\widetilde{q_{1,\textrm{nl}}}|_{x=0} = - \widetilde{q_3},
\label{eqn:disp_2nl22}
\end{align}
describes the dispersion relation for the waves localized at the interface between the nonlinear and linear layer. This equation has a structure that resembles  Eq.~(7) in Ref.~\cite{Agranovich80}. The differences between the two expressions result from different assumptions on the type of the nonlinearity used, as described at the beginning of section \ref{sec:nl-wv-eq}. The second solution,
\begin{align}
\widetilde{q_3} = - \widetilde{q_4},
\label{eqn:2llin}
\end{align}
gives the linear plasmon dispersion relation at the interface between two linear layers (3 and 4) \{compare with Eq. (2.12) in Ref.~\cite{Maier07}\}.

\subsubsection{Linear case}
\label{sec:fbm-lin}
In order to obtain the limiting expressions for the linear case ($\alpha_1~\rightarrow~0$) in the FBM one must assume that $x_0~\rightarrow~+\infty$. This can be understood by looking at the formula for the magnetic field shape in the nonlinear layer (\ref{eqn:H_nl}). For large values of $x_0$ the argument of the hyperbolic cosine tends to $-\infty$, i.e. in this case
 Eq.~(\ref{eqn:H_nl}) reduces to
\begin{align}
 H_{1}(x) &\propto e^{[k_0 q_1 (x - x_0)]}
\end{align}
This means that the field in the layer $1$ is now described by a decaying exponential which is in agreement with the solution of the Maxwell's equations in this layer in the linear regime.

Now the dispersion relation in the limiting case for three- and two-layer structures in the linear regime can be computed. Letting $x_0~\rightarrow~+\infty$, from Eq.~(\ref{eqn:q1nltilde}) one obtains that $\widetilde{q_{1,\textrm{nl}}}~\rightarrow~\widetilde{q_1}$. In this case, Eq.~(\ref{eqn:disp_3nl}) becomes
\begin{align}
\tanh(k_0 q_3 d) = - \frac{\widetilde{q_3}( \widetilde{q_{1}} + \widetilde{q_4})}
					 {\widetilde{q_3}^2 + \widetilde{q_4} \widetilde{q_{1}}}.
\label{eqn:disp_3l}
\end{align}
After some algebra, it transforms to
\begin{align}
e^{-2k_0 q_3 d} = \frac{(\widetilde{q_3} + \widetilde{q_1})(\widetilde{q_3} + \widetilde{q_4})}
					{(\widetilde{q_3} - \widetilde{q_1})(\widetilde{q_3} - \widetilde{q_4})},
\label{eqn:disp_3l2}
\end{align}
which is equivalent to Eq.~(2.28) in Ref.~\cite{Maier07} giving the dispersion relation for linear plasmons of a metallic film sandwiched between two linear dielectrics (insulator/metal/insulator structure --- IMI) or of a dielectric film sandwiched between two metals (metal/insulator/metal structure --- MIM).

For two-layer structure it is now straightforward to see that if  $\widetilde{q_{1,\textrm{nl}}}~\rightarrow~\widetilde{q_1}$ then Eq.~(\ref{eqn:disp_2nl22}) is reduced to the dispersion relation of the linear case [Eq.~(\ref{eqn:2llin})].

These three limiting cases show that our extended FBM fully recovers already known dispersion relations, including nonlinear ones, of simpler structures.

\subsection{Exact model}

In order to check the agreement between the results of our EM and the previously published results~\cite{Yin09} the limiting case of the EM nonlinear dispersion relation for the three-layer structure is considered.

Assuming that $L~\rightarrow~0$, Eq.~(\ref{eqn:phiYin}) simplifies to
\begin{equation}
\phi = \frac{\Psi^+_+ e^{ - k_0 q_3 d} + \Psi^-_- e^{ k_0 q_3 d}}
                {\Psi^+_- e^{  k_0 q_3 d} + \Psi^-_+ e^{ - k_0 q_3 d}}.
\label{eqn:phiYin3l}
\end{equation}
In the next step the expressions (${1}/{\phi} - 1$) and (${1}/{\phi} + 1$) appearing in Eq.~(\ref{eqn:Ez}) are expanded. Using Eqs.~(\ref{eqn:phiYin3l}) and (\ref{eqn:psi}), after lengthy but simple algebra one obtains
\begin{subequations}
\label{eqn:simpleYin3}
\begin{align}
\frac{1}{\phi} - 1 = 2M\bar{\epsilon}_3 [\bar{\epsilon}_3 \sinh(k_0 q_3 d) + 
									    \bar{\epsilon}_4 \cosh(k_0 q_3 d)],\\
\frac{1}{\phi} + 1 = 2M\bar{\epsilon}_3 [\bar{\epsilon}_3 \cosh(k_0 q_3 d) + 
									    \bar{\epsilon}_4 \sinh(k_0 q_3 d)],
\end{align}
\end{subequations} 
where $\bar{\epsilon}_k = {\epsilon_k}/{q_k}$ (for $k\in \{2,3,4\}$)
and
\begin{equation}
M = \frac{1}{(\bar{\epsilon}_2 - \bar{\epsilon}_3)(\bar{\epsilon}_3 + \bar{\epsilon}_4)e^{k_0 q_3 d} + 
		      (\bar{\epsilon}_2 + \bar{\epsilon}_3)(\bar{\epsilon}_3 - \bar{\epsilon}_4)e^{-k_0 q_3 d}}.
\end{equation}
As an intermediate step Eq.~(\ref{eqn:Ez}) is rewritten in the form:
\begin{equation}
E_{z,0}^2 = \frac{\left({\epsilon_2 \beta}/{q_2} \right)^2 \left({1}/{\phi} - 1  \right)^2}
             {\epsilon_{1,0}^2 \left({1}/{\phi} + 1 \right)^2 + \left({\epsilon_2 \beta}/{q_2} \right)^2 \left({1}/{\phi} - 1  \right)^2 } E_0^2.
\label{eqn:Ez20}
\end{equation}
Inserting Eqs.~(\ref{eqn:simpleYin3}) into Eq.~(\ref{eqn:Ez20}) and defining
\begin{subequations}
\begin{align}
Q = q_4 \epsilon_3 \tanh(k_0 q_3 d) + q_3\epsilon_4,\\
R = q_4 \epsilon_3 + q_3 \epsilon_4 \tanh(k_0 q_3 d)
\end{align}
\end{subequations} 
one obtains
\begin{equation}
E_{z,0}^2 = \frac{\beta^2\epsilon_3^2 Q^2 E_0^2}
			    {\beta^2 \epsilon_3^2  Q^2 +
                           \epsilon_1^2 q_3^2       R^2  }.
\label{eq:em-lim3}
\end{equation}
This last equation is identical to formula (11) in Ref.~\cite{Yin09}. Equation~(\ref{eq:em-lim3}) is then inserted into Eq.~(\ref{eqn:yin7}) in order to obtain the dispersion relation for a three-layer structure. The procedure of transforming this equation to obtain two separate dispersion relations, on a linear/nonlinear interface and a linear/linear interface ($d~\rightarrow~\infty$), is described in  Ref.~\cite{Yin09}.

\section{Results}
\label{sec:results}

As it was already mentioned above, theoretical studies of plasmon--solitons or more generally nonlinear localized surface waves started more than 30 years ago with the seminal paper of Agranovich \textit{et~al.}~\cite{Agranovich80}.
However, no experimental results confirming the existence of these nonlinear waves propagating in metal--dielectric structures have been provided. Consequently, from the modeling point of view, the main challenge is to design a feasible structure, that enables the experimental realization of plasmon--soliton coupling. 

To reach this goal, several conditions must be satisfied simultaneously. Firstly, a structure that supports plasmon--solitons of a solitonic type (with a pronounced soliton peak inside nonlinear dielectric which facilitates experimentally both its excitation and its discrimination from linear waves) must be found. Secondly, solutions should appear for physically realistic combinations of material parameters, beam power, and nonlinear coefficient. The last, more practical and supplementary requirement is to design a structure in which the plasmon field is accessible both for measurements using the tip of a Scanning Near-Field Optical Microscope and for potential applications such as sensing~\cite{Taya08,Jha09,Dinesh11,Homola06}. 

This task has already been fulfilled in Ref.~\cite{Walasik12} in which a simple structure that supports low-power plasmon--solitons is described. Nevertheless, not all the details of the design process were given there. They are provided in this section, that gives also a complete description of all the nonlinear stationary solutions that can be generated in planar structures made of a combination of semi-infinite nonlinear dielectric, metal film, and linear dielectric layers. This section starts with the two-layer configuration and finishes with the four-layer one which is shown to be the simplest device that fulfills all the requirements to facilitate the experimental observation of plasmon--solitons as defined above. 

\subsection{Two-layer configuration}

In the case of the two-layer configuration (single interface between a nonlinear dielectric and a metal) the only nonlinear solutions that we are able to find using our three models are of the plasmonic type (no pronounced soliton peak in the nonlinear medium). This results are in agreement with the conclusions drawn by looking at the solutions of the FBM and the continuity conditions for the field at the interface. The main results from the FBM are summarized:
\begin{itemize}
\item the field in the nonlinear material is described by the formula (\ref{eqn:H_nl}) with the free parameter $x_0$,
\item the field in the metal is given by the exponential function (\ref{eqn:H_4}) (with $L=d=0$) and decreases to 0 as $x$ tends to infinity to satisfy the boundary condition $H_y \xrightarrow[]{x\rightarrow + \infty} 0$,
\item in order to obtain the nonlinear dispersion relation we use the conditions for the continuity of the fields at the interface $x=0$:
\begin{enumerate}
\item for the magnetic field $H_{1} = H_{4}$, so that in Eq.~(\ref{eqn:H_4}) $C=H_0$, \label{cond1}
\item for the longitudinal component of the electric field $E_{z,1} = E_{z,4}$, which using Eq.~(\ref{eqn:rotH42}) is expressed in terms of the $x$-derivative of $H_y$ and the permittivity of the media:
\begin{equation}
\frac{1}{\epsilon_1} \frac{d H_{1}}{dx}=\frac{1}{\epsilon_4} \frac{d H_{4}}{dx}.
\end{equation} 
\end{enumerate}
\end{itemize}
Because the permittivities of the metal and the nonlinear dielectric have opposite signs ($\epsilon_1\epsilon_4<0$), from the condition B it is seen that the derivatives of $H_y$ must have opposite signs at both sides of the interface. 
From Eq.~(\ref{eqn:H_4}) it follows that $\left. \frac{d H_{4}}{dx}\right|_{x=0^+} < 0$. This implies that the derivative on the nonlinear side of the interface has to be positive $\left( \left. \frac{d H_{1}}{dx}\right|_{x=0^-} > 0\right)$. By looking at Eq.~(\ref{eqn:H_nl}) one can see that this condition is fulfilled only if $x_0>0$. This allows us to conclude that only the plasmonic type solutions exist on a single metal/nonlinear dielectric interface.

\subsection{Three-layer configuration}

In this section results obtained for three-layer configurations ($L$ is set to 0) are presented. First, to confirm the validity of our FBM its results are compared with the results from Ref.~\cite{Ariyasu85}. Then the general classification of nonlinear solution types is described and illustrated. Finally the structure parameter scans are performed in order to find configurations supporting low power plasmon--solitons. 

\subsubsection{Comparison of FBM results with those of Ariyasu et al.~\cite{Ariyasu85}}
\label{sec:ari_comp}

In section~\ref{sec:ver-fbm}, it was shown that the nonlinear dispersion relation for the four-layer FBM reproduces several known analytical results including these for the three-layer model proposed in Ref.~\cite{Ariyasu85}. In order to check the validity of our model, the graphical comparisons between the nonlinear dispersion curves for the three-layer structure presented in Ref.~\cite{Ariyasu85} and the results of our modeling are presented. The parameters used in our simulations are the same as those used in Fig.~1 of Ref.~\cite{Ariyasu85}. The linear part of the nonlinear medium refractive index is $\tilde{\epsilon}_{l,1} = 16-0.0096i$, metal permittivity is $\tilde{\epsilon}_3 = -1000 -160i$, and the linear dielectric permittivity is $\tilde{\epsilon}_4 = 16$. The thickness of the metal film is $d = 50$~nm, the wavelength used is $\lambda = 5.5$~$\mu$m, and the nonlinear parameter is $n_2^{(1)} = 10^{-7}$~m$^2$/W.

\begin{figure}[!ht]
\centering
\includegraphics[width = 0.238\textwidth,angle=-0,clip=true,trim= 55 40 205 8]{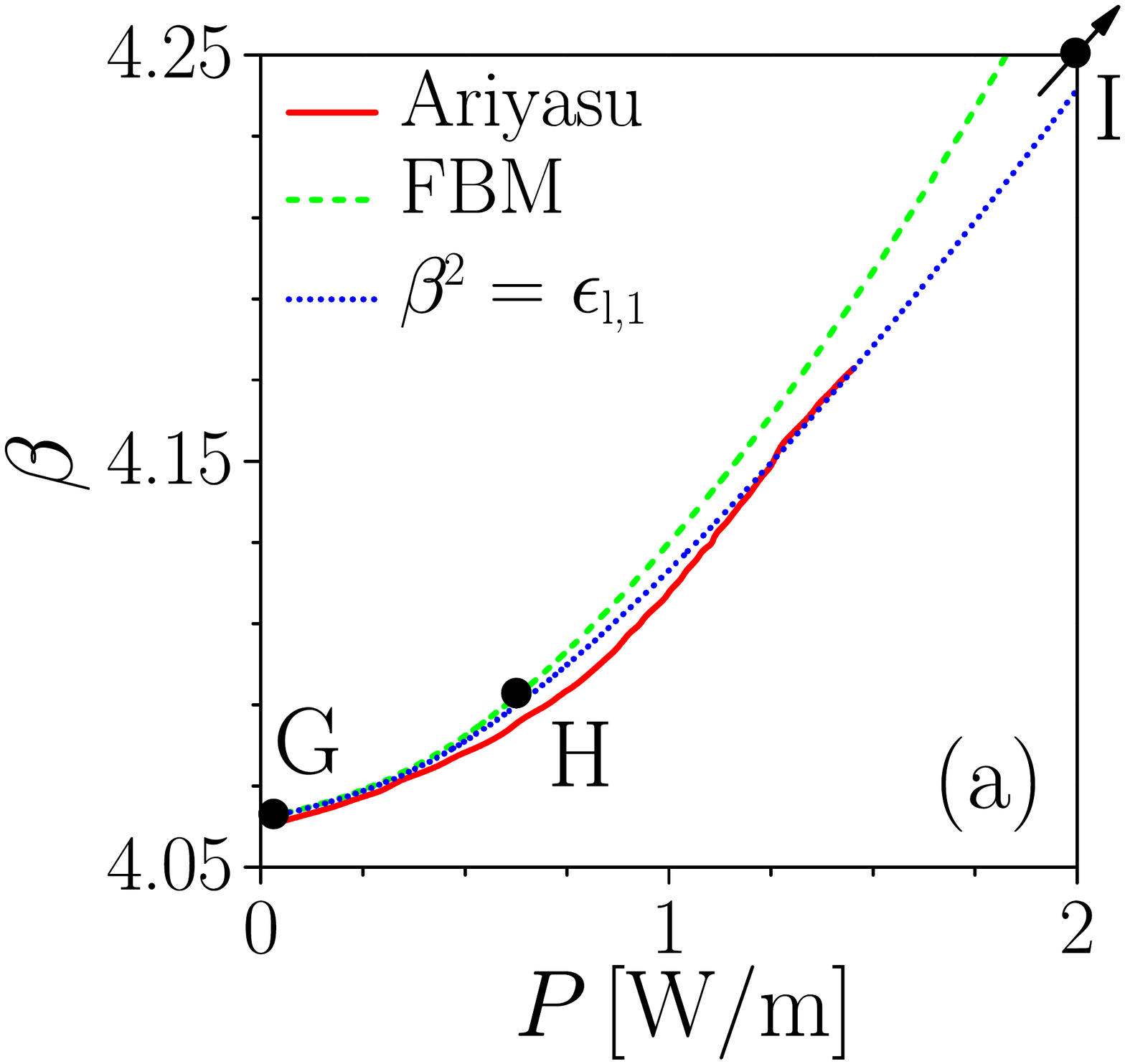}
\includegraphics[width = 0.238\textwidth,angle=-0,clip=true,trim= 55 40 205 8]{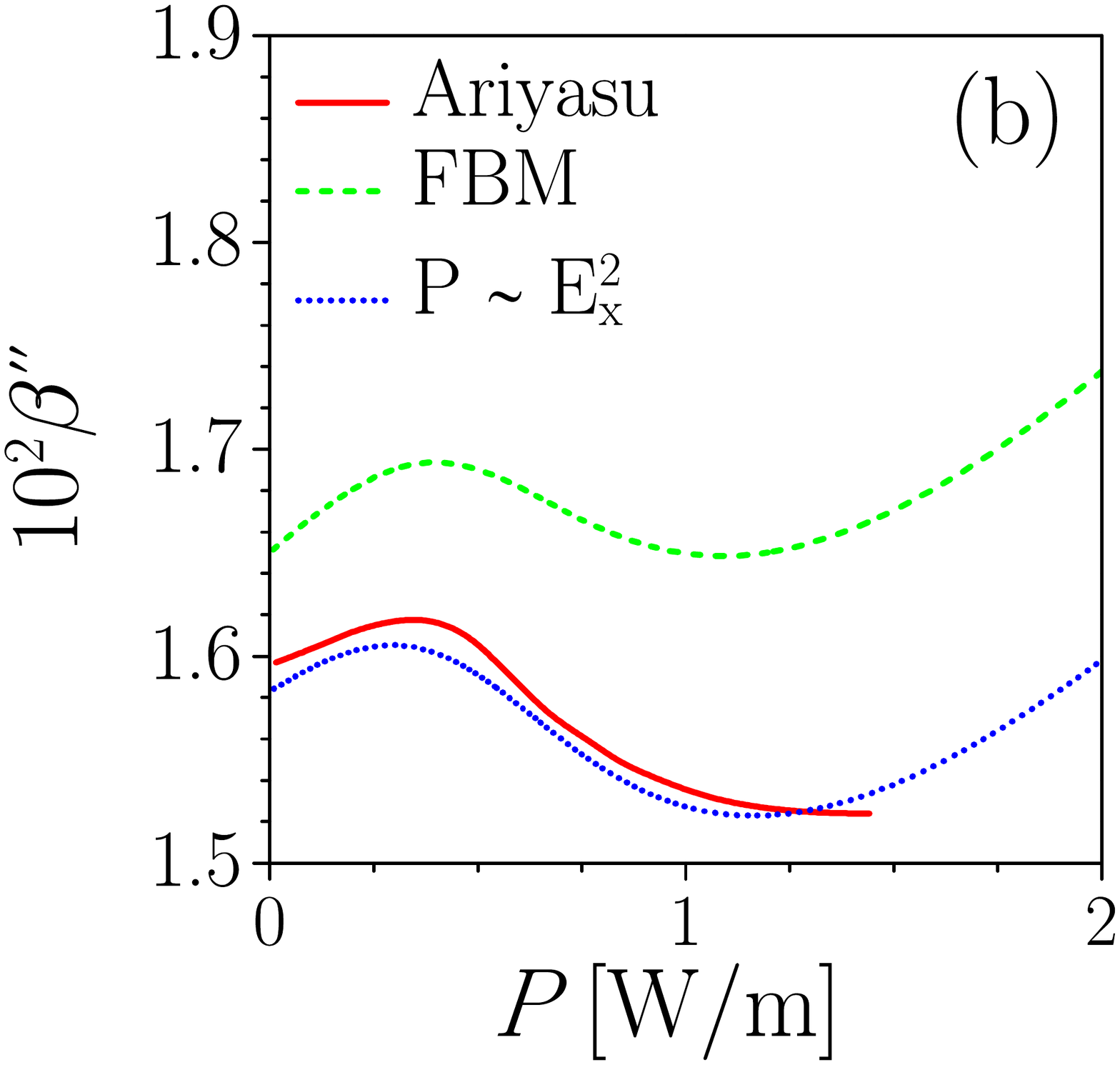}\\
\includegraphics[width = 0.238\textwidth,angle=-0,clip=true,trim= 15 40 245 8]{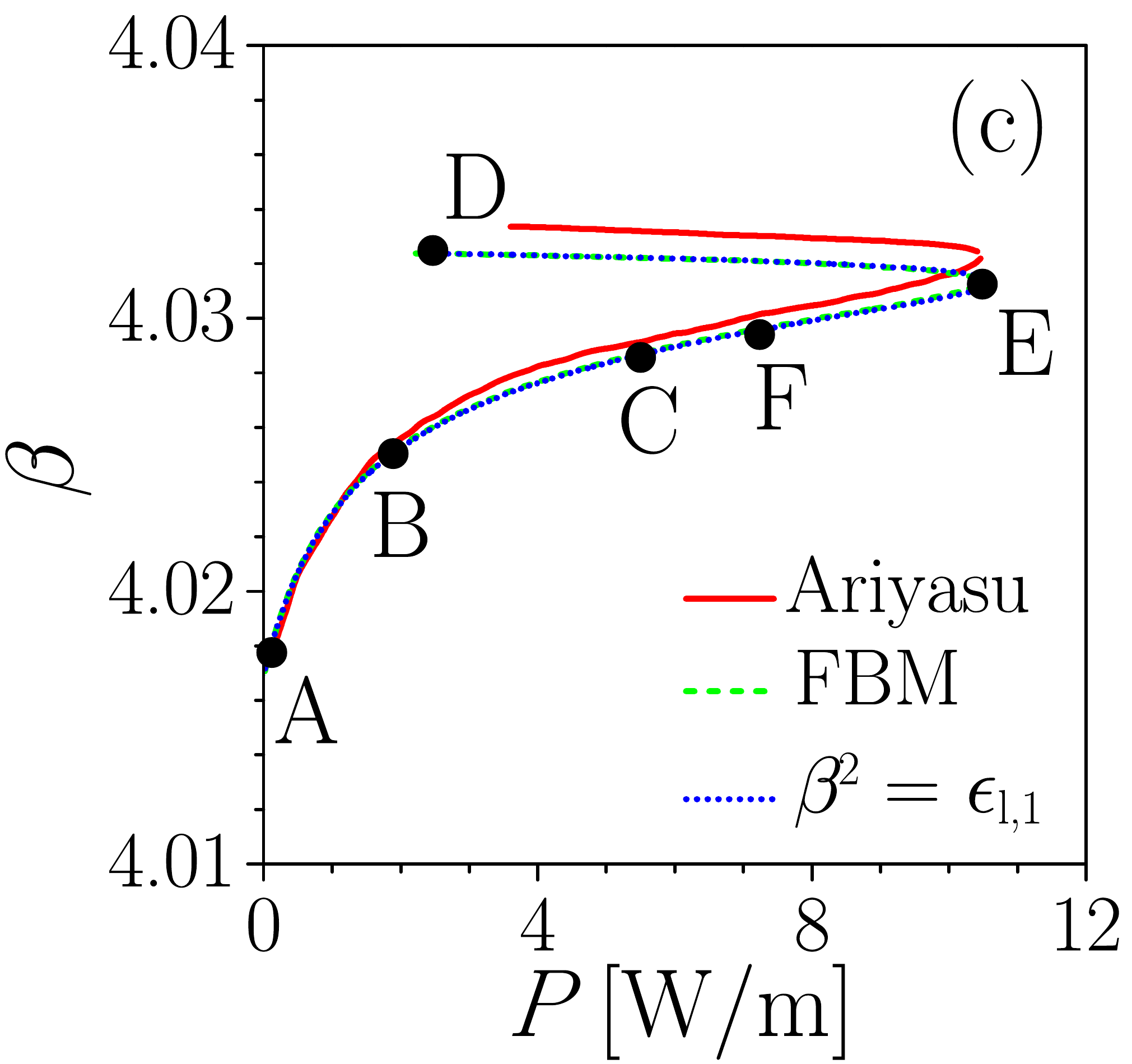}
\includegraphics[width = 0.238\textwidth,angle=-0,clip=true,trim= 70 40 190 8]{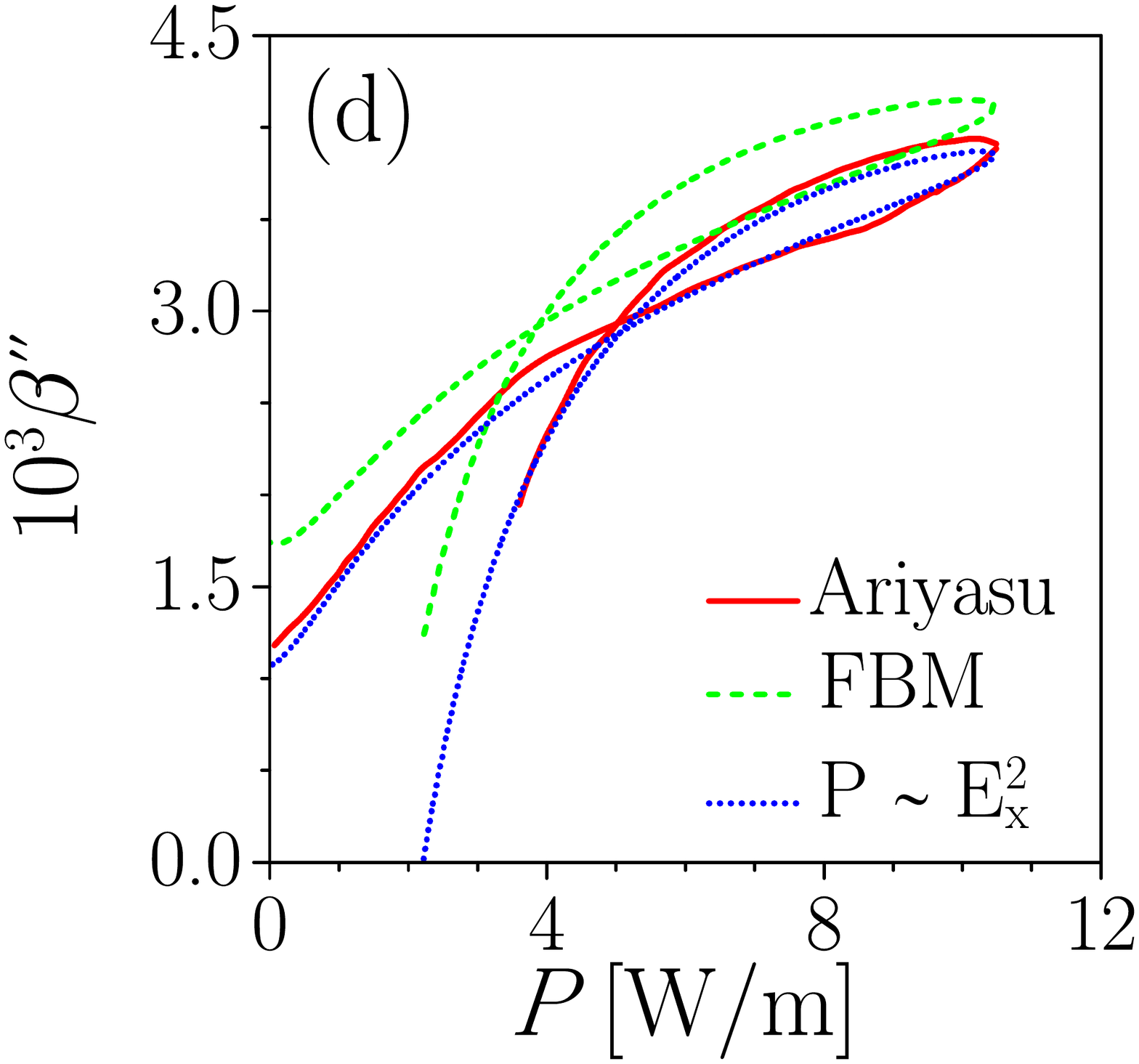}
\caption{Comparison of the original results from the article of Ariyasu \textit{et~al.}~\cite{Ariyasu85} (Fig.~1 digitized) (red solid curve) and results obtained with our FBM (green dashed curve) with some specific approximations (blue dotted curve) (see the text for more details). Real part [(a), (c)] and imaginary part [(b), (d)] of the dispersion relation for the three-layer structure. In panel (c) the green and blue curves overlap perfectly. The labelled points A--I correspond to the field shapes depicted in Fig.~\ref{types_profiles}. Point I lays out of the plotting range (see section~\ref{sec:mod_class} for explanation).}
\label{fig:ari-comp}
\end{figure}

Figures~\ref{fig:ari-comp}(a) and (c) show the dispersion relation in which the real part of the effective index $\beta$ is plotted as a function of the power density of the nonlinear wave $P$. The original results from Ref.~\cite{Ariyasu85} are depicted by the red solid curve and the results obtained with our FBM for the three-layer structure are presented by the green dashed curve. For the low effective index branch the two curves are in relatively good agreement. On the other hand, for the high effective index branch small discrepancy between the results appears. 

Two reasons explain the differences between these curves. Firstly, a different form of the nonlinear permittivity tensor is used (see section \ref{sec:nl-wv-eq}) and as a consequence different values of the effective nonlinear function $a(x)$ in Eq.~(\ref{eqn:wave4}) (compare with $\alpha'$ defined in Eq.~(4b) in Ref.~\cite{Ariyasu85}). The difference between the two solutions is small for the parameter range, where the effective index is close to the linear refractive index of the nonlinear material and larger for higher values of the effective index. This is in full agreement with the explanation presented at the end of the section \ref{sec:nl-wv-eq}.
Secondly, a closer examination of Eq.~(4b) and Eqs.~(9)--(11) in Ref.~\cite{Ariyasu85} reveals that to compute the power the authors made the approximation $\beta^2 = \epsilon_{l,1}$. 
To reproduce the original results provided in Fig.~1 of Ref.~\cite{Ariyasu85} this approximation for power calculations is used in our model for the test purpose. The corresponding blue dotted curve on Fig.~\ref{fig:ari-comp}(a) is closer to the original results than the green curve obtained with our full FBM.

Figures~\ref{fig:ari-comp}(b) and (d) show the comparison of the original results from Ref.~\cite{Ariyasu85} and our results for the dependency of the imaginary part of the effective index $\beta''$ as a function of the power density.  The results obtained with our FBM (green dashed curve) lay slightly above the original results (red solid curve). The comparison of the formulas used to calculate losses \{Eq.~(8) in Ref.~\cite{Stegeman85} and  Eq.~(\ref{eqn:Beta_im1d_fin}) for our formulation\} shows that losses are calculated in different ways. In Ref.~\cite{Ariyasu85} authors use Eq.~(8) from Ref.~\cite{Stegeman85}, where losses are proportional to the product of the imaginary part of permittivity with the power density $P$ in each layer ($\beta'' \propto \int \epsilon'' P dx$). The power density is proportional to the Pointing vector and in the frame of a linear approximation $P \propto E^2_x$. In our formulation [Eq.~(\ref{eqn:Beta_im1d_fin})] the losses [green curve in Figs.~\ref{fig:ari-comp}(b) and (d)] depend on both components of the electric fields [$\beta'' \propto \int \epsilon'' (E_x^2 + E_z^2) dx$]. If a formulation in which the losses are proportional only to the transverse field component in our FBM is used, a very good agreement with the original results is reached [see the blue dotted curve in Figs.~\ref{fig:ari-comp}(b) and (d)].

Even if small numerical discrepancies between our improved approach and the original results of Ariyasu \textit{et~al.} appear due to different approximations used, they are fully understood. Our extended FBM is able to reproduce the results published by Ariyasu \textit{et~al.} with good agreement.

\subsubsection{Nonlinear wave types classification}
\label{sec:mod_class}

\begin{widetext}

\begin{SCfigure}[][!hb]
\centering
\includegraphics[width = 0.75\textwidth,angle=-0,clip=true,trim= 123 385 119 52]{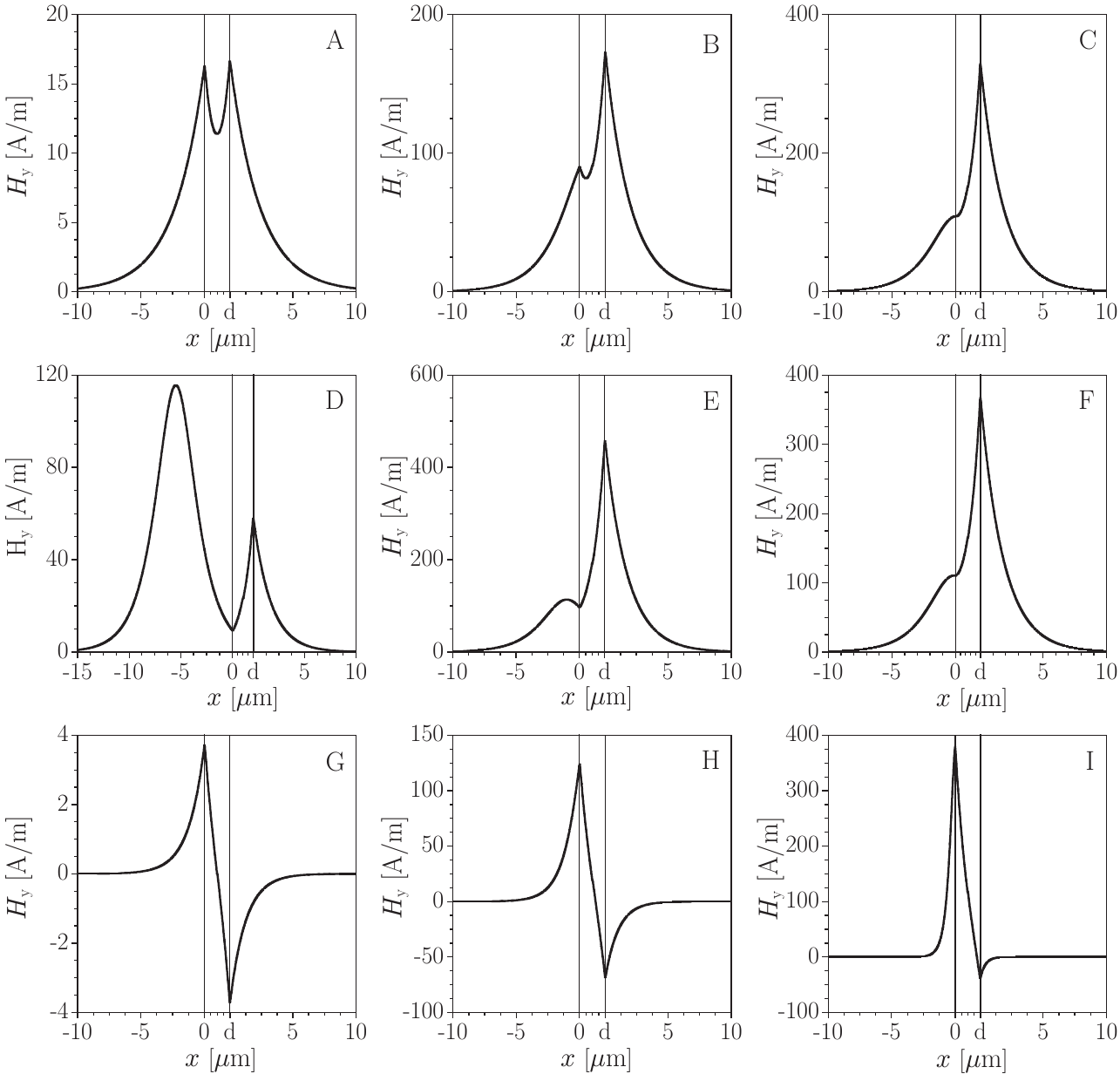}
\caption{Magnetic field component $H_y$ profiles for the three-layer structure described in section \ref{sec:ari_comp} corresponding to the points indicated on the dispersion plot in Fig.~\ref{fig:ari-comp}. In all the figures showing field shapes in this paper the coordinates inside the thin intermediate films are not at the same scale as those used in the other layers for a better visibility of the field behavior. In the first row the symmetric type nonlinear plasmons are shown. In the second row the nonlinear plasmon--solitons are shown and in the last row the antisymmetric type nonlinear plasmons are shown. The columns correspond to different values of $|x_0|$: the first column to 5.5~$\mu$m, the second column to 1~$\mu$m, the third column to 0.1~$\mu$m.}
\label{types_profiles}
\end{SCfigure}

\end{widetext}

In this section a classification of the types of solutions that exist in the three-layer structures is presented. It is useful for the remaining part of this work to classify and name different types of solutions as they will be similar in four-layer configurations. In Fig.~\ref{fig:ari-comp} nine points were labeled from A to I in order to describe the type and the transformation of solutions along the nonlinear dispersion curve. The magnetic field profiles corresponding to these points are shown in Figs.~\ref{types_profiles}A--I.

From the analytical considerations it has already been seen in section \ref{sec:fbm-lin} that for $x_0\rightarrow+\infty$ the solutions correspond to the linear limiting case. In this case for the symmetric three-layer IMI structure two solutions exist: symmetric (long range) plasmon and antisymmetric (short range) plasmon~\cite{Homola06,Berini09}. Points A and G were obtained for $x_0 = \lambda = 5.5$~$\mu$m and the corresponding solutions are close to the linear ones. For both solutions the power density is relatively low $P < 0.1$~W/m (this type of solution is obtained for even lower powers if one selects larger values of $x_0$).
The corresponding field shapes are like the linear solutions. Figure \ref{types_profiles}A presents a field shape that is very close to the symmetric linear plasmon and Fig.~\ref{types_profiles}G shows a field profile very similar to the antisymmetric linear plasmon.

In the following the field transformation along the dispersion curves is described in details. First, the transformation of the symmetric type plasmonic solutions, located at the lower branch of the dispersion curve, is studied. Decreasing the value of $x_0$ to 1~$\mu$m (all other parameters being identical) we obtain the field shape corresponding to the point B. The power density of this nonlinear wave is $P \approx 2$~W/m and the field shape still resembles the symmetric linear plasmon but the field is now asymmetric and the energy is more localized on the interface between the metal film and the linear dielectric. Upon further decrease of the value of $x_0$ to 0.1~$\mu$m (point~C) the power density of the solution increases to $\approx5.5$~W/m and the field shape becomes even more asymmetric. 
The solutions described above are referred as symmetric type nonlinear plasmons.

When $x_0$ becomes negative one obtains a new class of solutions, where the local magnetic field maxima are located both at the interface between the metal film and the linear dielectric and inside the nonlinear medium.
Upon decrease of the $x_0$ parameter down to $-0.1$~$\mu$m the power density still increases (to around $7.5$~W/m corresponding to point F) and reaches its maximum at the point E for $x_0=-1$~$\mu$m. Further reduction of $x_0$ leads to the decrease of the total power density ($P \approx 2.5$~W/m for point D corresponding to $x_0 = -5.5$~$\mu$m). Point D lays close to the end of the branch corresponding to $x_0 \rightarrow -\infty$ associated with the isolated classical soliton that does not interact with the metal film. Even though the field profiles C and F at the first glance look almost identical, there is an important qualitative difference between them. On one hand profile C ($x_0 = 0.1$~$\mu m$) is classified as plasmonic type solution because there is no field maximum in the nonlinear layer. On the other hand, profile F ($x_0 = -0.1$~$\mu m$) does have a local maximum in the nonlinear layer (located close to the metal interface) and therefore it belongs to another class of solutions. 

For all the solutions presented in Figs.~\ref{types_profiles}D--F the peak amplitude of the solitonic part (in the nonlinear dielectric) remains at approximately the same level and only the maximum of the plasmon field on the metal/linear dielectric interface decreases with the decrease of $x_0$ value. All the solutions in the second row of Fig.~\ref{types_profiles} will be called solitonic type solutions or nonlinear plasmon--solitons. 

It is also worth noting that the solitonic type solution can not be obtained at any desired power density. Following the dashed green curve in Fig.~\ref{fig:ari-comp}(c) and knowing the field shapes one can see that for power densities between $6.5$~W/m and $10.5$~W/m two solitonic type solutions with different $x_0$ correspond to one power density. For power densities between $2.5$~W/m and $6.5$~W/m and for a maximum power density of $10.5$~W/m there is only one solitonic type solution corresponding to each power density. Below $2.5$~W/m and above 10.5~W/m no solitonic type solution exists.

Finally, the transformation of solutions laying along the upper branch of the dispersion relation [see Fig.~\ref{fig:ari-comp}(a)] is described. The branch starts with the solution described above, very similar to the antisymmetric linear plasmon (point G). Decreasing the value of $x_0$ to 1~$\mu$m results in the field profile corresponding to point H. The field shape of this solution is like the antisymmetric linear solution but it is distorted. The field distribution is asymmetric and this time the field is more localized at the metal/nonlinear dielectric interface (contrarily to the case of symmetric type solutions, where the field tends to localize on the opposite metal interface). Decreasing $x_0$ even further down to 0.1~$\mu$m we obtain the field shape presented in Fig.~\ref {types_profiles}I. Here the field is almost entirely localized at the metal/nonlinear dielectric interface and is therefore even more asymmetric. The corresponding power density is 2.5~W/m and the effective index is so high ($\beta = 4.57$) that it is outside of the plot in Fig.~\ref{fig:ari-comp}(a). The solutions presented in Figs.~\ref{types_profiles}G--I will be called antisymmetric type nonlinear plasmons.

\subsubsection{Low-power solution search}
\label{sec:3-lay-low-search}
The simplest structures in which it is possible to obtain the solutions of the plasmon--solitons type are three-layer structures, as it has already been shown in section \ref{sec:mod_class}.

The study presented in Ref.~\cite{Ariyasu85} deals only with configurations, where the linear parts of the permittivities of linear and nonlinear dielectrics are equal. Below a more general case is studied, in which a permittivity contrast between the linear and nonlinear dielectric is introduced. For this study the FBM limited to three-layers ($L$ is set to 0 and only layers 1, 3 and 4 are present) is used. The configurations where $\epsilon_{l,1} \ge \epsilon_4$ are chosen to guarantee that the solutions are localized at the interface between layers 3 and 4 as  $\beta \ge \sqrt{\epsilon_{l,1}}$ [see Eqs.~(\ref{eqn:q-def}) and (\ref{eqn:H_4})]. From the practical point of view this condition can also be justified by looking at typical material properties. For the glasses it is known that, in most cases, the nonlinear coefficient $n_2$ increases with the increase of the linear refractive index~\cite{Zakery07,Boudebs01}. This justifies our choice to consider a linear permittivity of the linear layer to be lower that of the nonlinear layer. 

\begin{figure}[!ht]
  \centering
\includegraphics[width = 0.231\textwidth,angle=-0,clip=true,trim= 31 43 180 15]{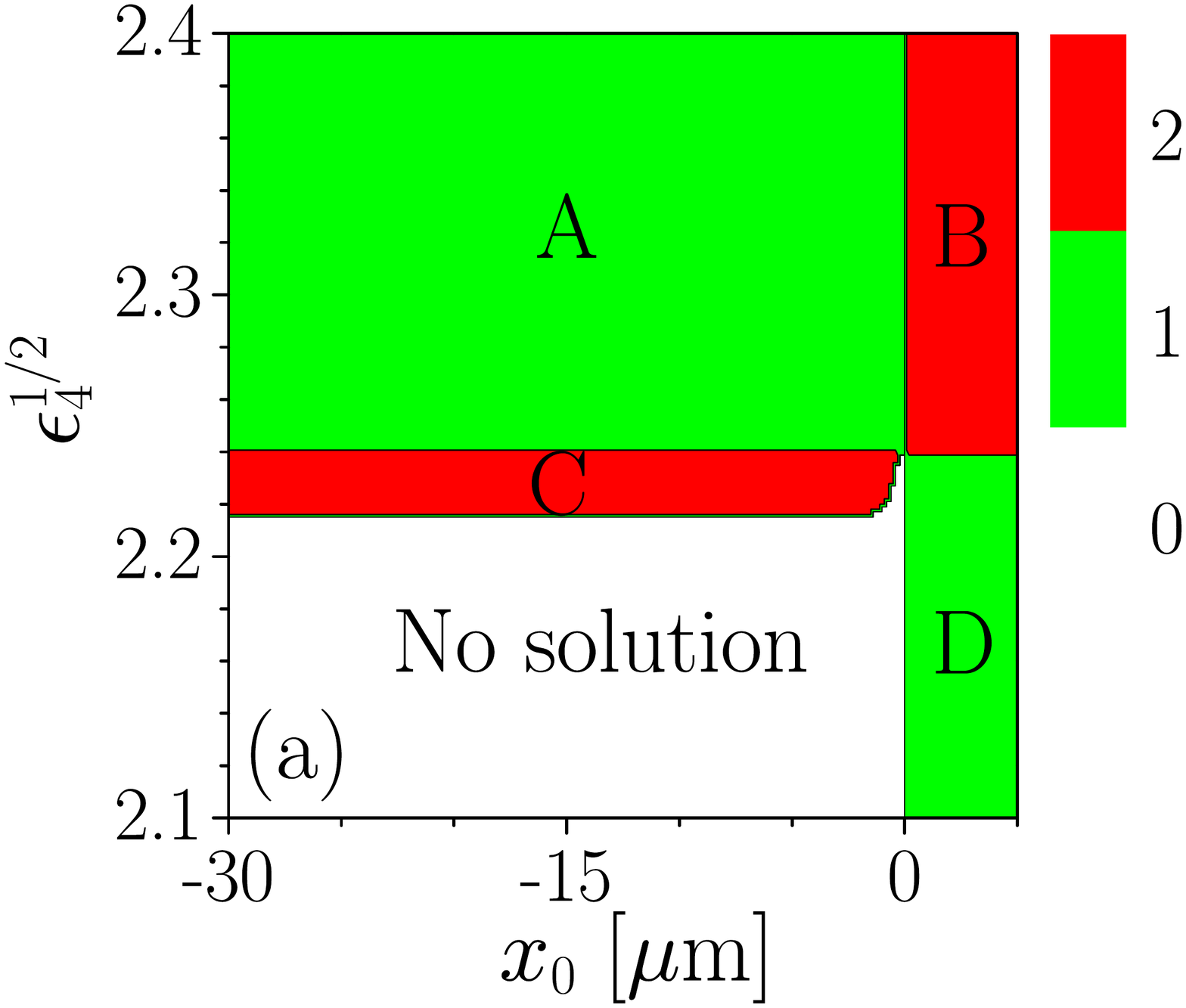}
\includegraphics[width = 0.245\textwidth,angle=-0,clip=true,trim= 12 42 160 15]{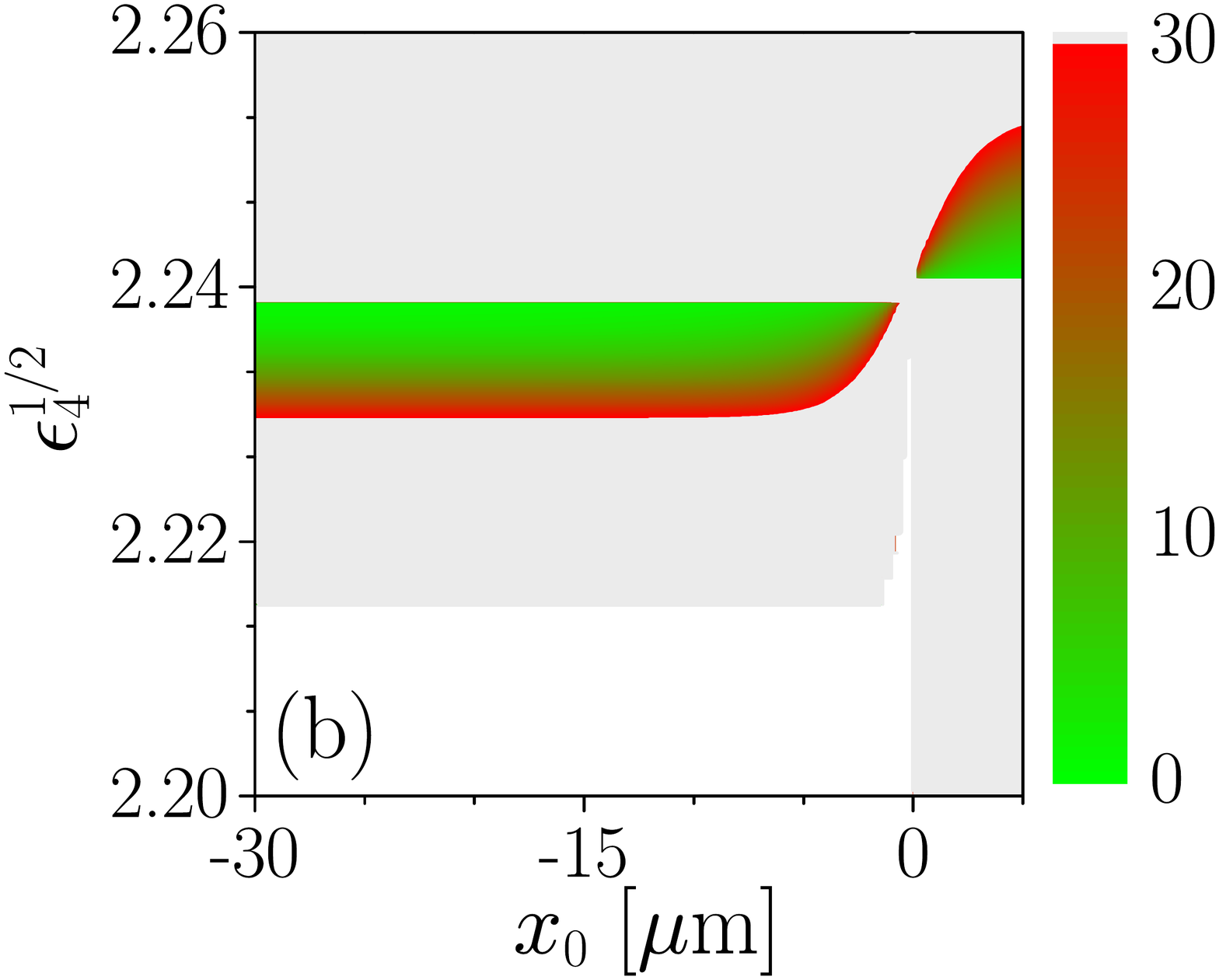}
\caption{(a) Number of solutions as a function of $x_0$ parameter and of the external linear layer refractive index ($\sqrt{\epsilon_4}$). (b)~Peak power [GW/cm$^2$]  for the low-power solutions close to the cut-off value of $\sqrt{\epsilon_4}$. In this and all the following peak power color maps in this paper, only solutions with peak power below 30 GW/cm$^2$ are plotted, the existence of solutions with higher peak power is marked with the gray color, and white denotes regions with no solutions. The parameters: $\epsilon_{l,1} = 2.4^2$, $n_2^{(1)} = 10^{-17}~\rm{m^2/W}$, $d = 40$~nm, $\epsilon_3 = -20$, and $\lambda = 1.55$~$\mu$m were used.}
\label{3_lay_nb_power}
\end{figure}

In order to obtain color maps in this section and the next one, the scans over parameters were performed using the FBM in such a way, that only solutions with the effective index  $ \sqrt{\epsilon_{l,1}} < \beta <4 \sqrt{\epsilon_{l,1}}$ were sought. For lower effective indices no localized solution exists as pointed out at the end of section \ref{sec:disp_rel_FBM} and higher effective indices are not interesting for our purpose because the corresponding solutions have extremely high power density and the nonlinear index modification is too high to be physically meaningful.

Figure \ref{3_lay_nb_power}(a) shows the dependency of the total number of solutions on the parameter $x_0$ and on the linear external dielectric refractive index $\sqrt{\epsilon_4}$.
For the symmetric structure ($\sqrt{\epsilon_4} =\sqrt{\epsilon_{l,1}} = 2.4$) (as discussed in section \ref{sec:mod_class}) and for quasi-symmetric configurations with low refractive index contrast $\Delta \epsilon = \epsilon_{l,1}-\epsilon_4 \lesssim 0.16$ one solitonic type solution (region A) and two (symmetric type and antisymmetric type) plasmonic solutions (region B) exist. Upon decrease of the linear layer refractive index (increasing the index contrast between nonlinear and external dielectrics) a narrow region (C) with two solitonic type solutions appears. These solutions do not exist for negative values of $x_0$ close to zero. Further decrease of the linear layer refractive index causes both solitonic type solutions to vanish around $\sqrt{\epsilon_4} = 2.22$. In the case of plasmonic type solutions ($x_0>0$) the decrease of the linear layer refractive index causes symmetric type solution to vanish (at a  cut-off index value of $\sqrt{\epsilon_4} \approx 2.24$) and only the antisymmetric type solution remains (region D) (even for $\sqrt{\epsilon_4} =1$ which is not shown on this plot).

Figure \ref{3_lay_nb_power}(b) shows the peak power of the solutions in a transition region close to the cut-off linear layer refractive index. The maximal peak power was set to $30 \rm{\;GW/cm^2}$ what, taking into account the nonlinearity parameter used $n_2^{(1)} = 10^{-17}~\rm{m^2/W}$, involves a maximum nonlinear index modification $\Delta n \le3\times 10^{-3}$. This value of $n_2$ is typical for chalcogenide glasses~\cite{Zakery07,Fatome09} or for hydrogenated amorphous silicon which seems to be a promising material for nonlinear integrated optics~\cite{Lacava13,Matres12}. It can be seen that the low-power solutions exist only in a very narrow range of $\sqrt{\epsilon_4}$ values. The solitonic type solutions have their lowest peak intensities slightly below the cut-off index and plasmonic type solutions above this value as it can be seen in Fig.~\ref{3_lay_nb_power}(b).

These studies confirm [for symmetric configuration  ($\epsilon_4 = \epsilon_{l,1}$)] and complete the results given in Ref.~\cite{Ariyasu85} (see line 5 of table I therein) to include the more general case ($\epsilon_4 \neq \epsilon_{l,1}$).

\begin{figure}[!ht]
\includegraphics[width = 0.238\textwidth,angle=-0,clip=true,trim= 7 22 75 2]{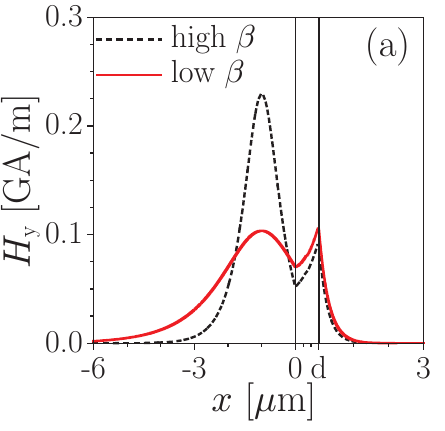}
\includegraphics[width = 0.238\textwidth,angle=-0,clip=true,trim= 5 22 77 2]{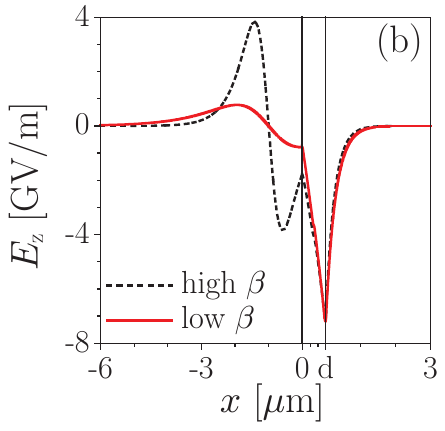}
\caption{Comparison of the field profiles $H_y(x)$ (a) and $E_z(x)$ (b) for the two plasmon--solitons existing in region C in Fig.~\ref{3_lay_nb_power}(a) for the same $x_0$ value.}
\label{2_sol_sol}
\end{figure}

Figure \ref{2_sol_sol} shows the comparison of the magnetic field $H_y$ and the longitudinal component of the electric field $E_z$ for the solitonic type solutions that appear in the three-layer structure for the same value of $x_0$ [region C in Fig.~\ref{3_lay_nb_power}(a)]. Here the parameters are $\epsilon_4 = 2.23^2$ and $x_0 = - 1$~$\mu$m. The solution with the lower effective index $\beta$ has a lower peak amplitude of the solitonic part than the one of the higher effective index solution. The solitonic part is broader and the plasmonic part peak amplitude is slightly higher in the former case.

\begin{figure}[!hb]
  \centering
\includegraphics[width = 0.235\textwidth,angle=-0,clip=true,trim= 45 45 175 10]{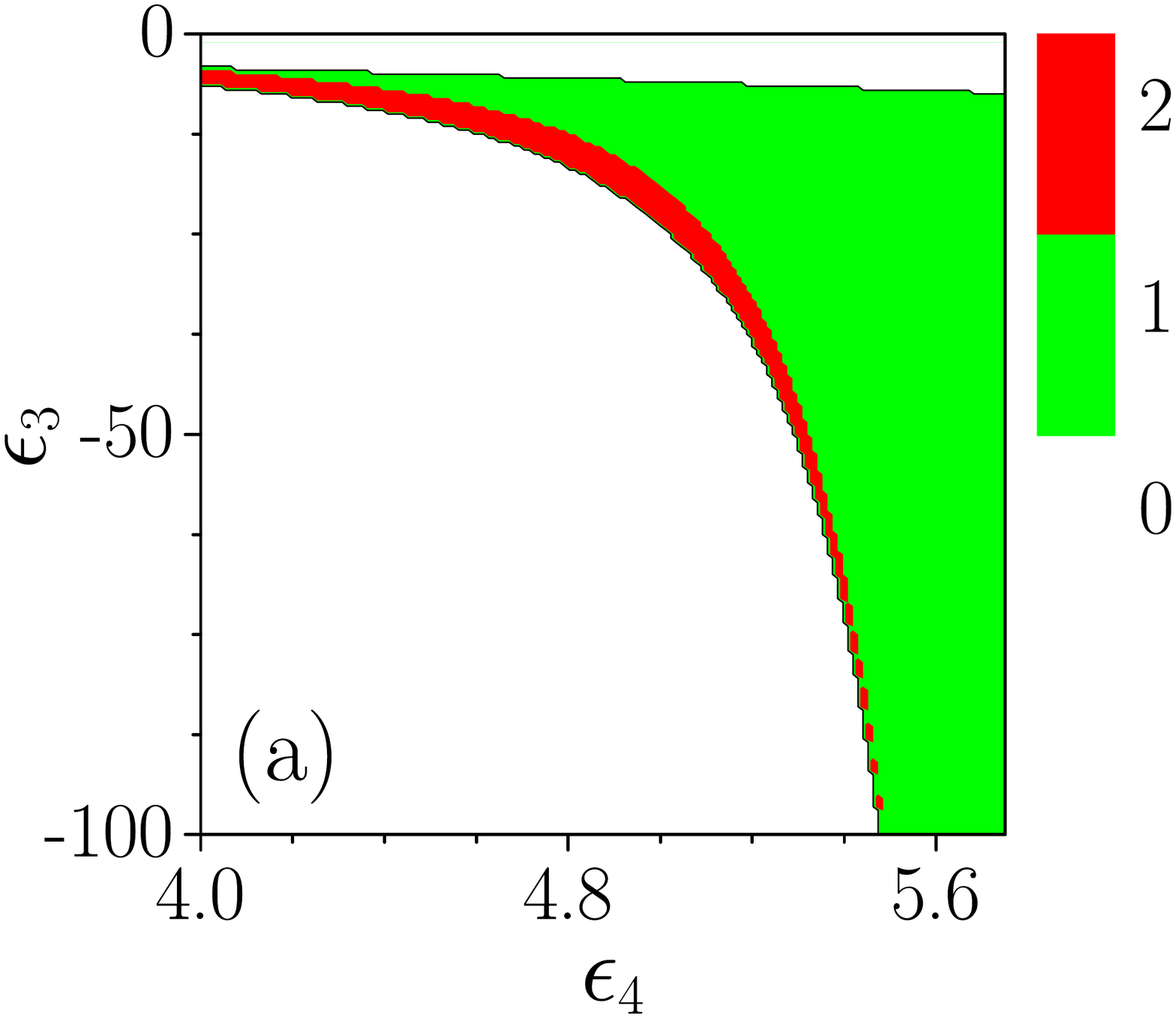}
\includegraphics[width = 0.241\textwidth,angle=-0,clip=true,trim= 45 45 155 10]{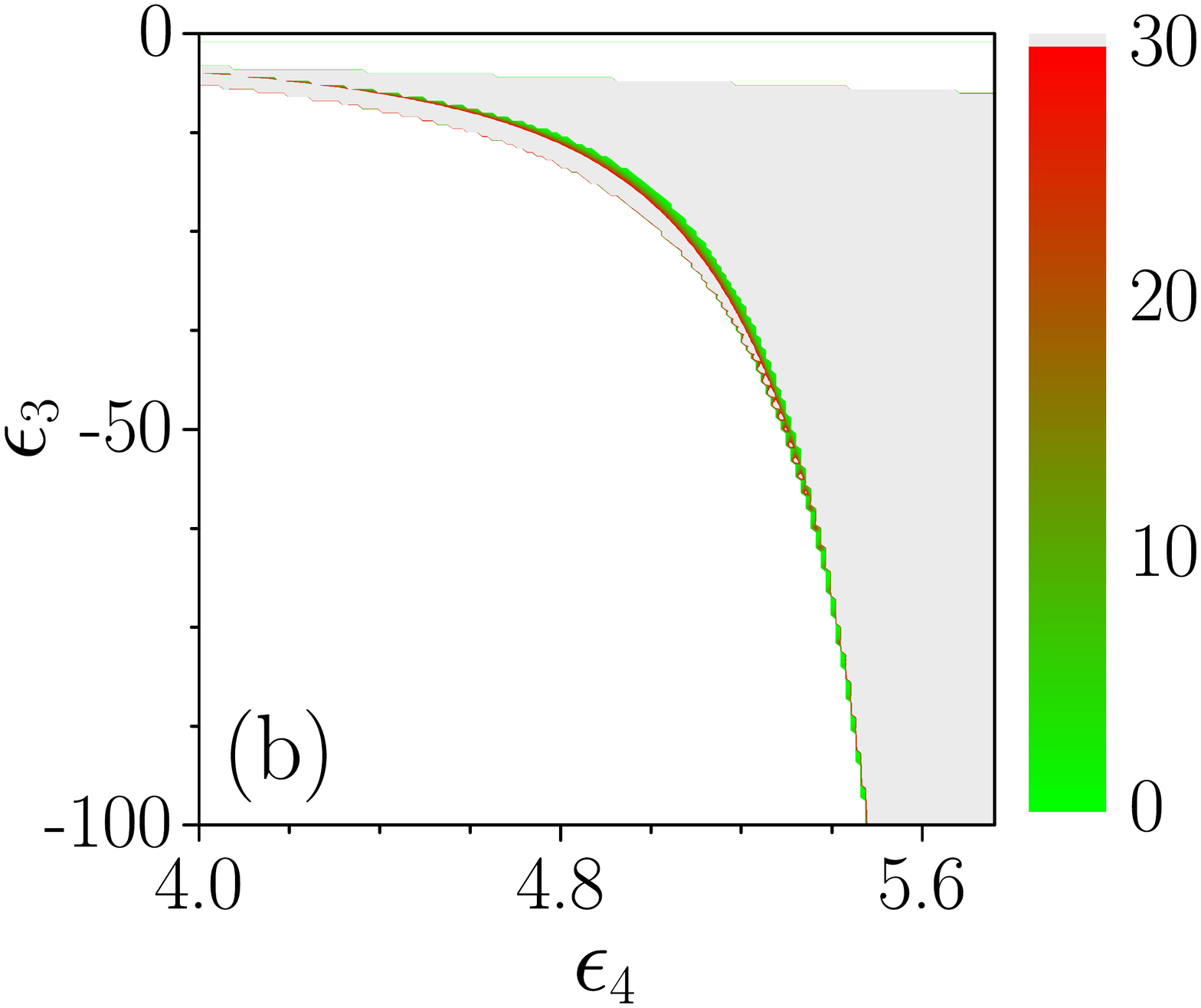}
\caption{(a) Number of solitonic type solutions in a three-layer structure with the same parameters as in Fig.~\ref{3_lay_nb_power} but for a fixed $x_0 = -15.5$~$\mu$m, as a function of the metal permittivity $\epsilon_3$ and of the linear dielectric permittivity $\epsilon_4$. (b)~Peak power [GW/cm$^2$] for the low-power solutions.}
\label{3_lay_nb_met}
\end{figure}

Now the influence of the metal permittivity changes on the behavior of the solitonic type solutions in three-layer structures is analyzed. The center of the solitonic part is set to be at the distance of 10 wavelengths from the metal film ($x_0 = -15.5$~$\mu$m). The number of solutions as a function of the metal permittivity and of the linear dielectric permittivity is studied. From Fig.~\ref{3_lay_nb_met}(a) it can be seen that two effects occur with the increase (decrease of the absolute value) of the metal permittivity. Firstly, the index contrast between layers 1 and 4 for which solutions can be found increases. Secondly, the allowed external dielectric permittivity range where two solitonic type solutions occur for one value of $x_0$ expands. There is also a cut-off metal permittivity above which no solution exists. This cut-off occurs when $|\epsilon_3| \approx \epsilon_{4}$. From Fig.~\ref{3_lay_nb_met}(b), that shows the peak power for low-power plasmon--solitons, it can be seen that the low-power solutions lay in a very narrow region close to the line separating regions with one and two solutions.

As a conclusion, we see that asymmetric structures (with $\epsilon_{l,1} > \epsilon_4$) are able to support the solitonic type solutions at much lower powers than symmetric structures. However, in order to obtain really low power densities the index contrast between the two dielectrics has to be precisely chosen [see Figs.~\ref{3_lay_nb_power}(b) and \ref{3_lay_nb_met}(b)]. The asymmetric three-layer configurations fulfill two out of three conditions set at the beginning of this section: they support both plasmonic and solitonic type plasmon--solitons and it is possible to obtain low-power solitonic type solutions. However, these solutions are obtained for configurations in which the linear medium refractive index is close to the linear part of the nonlinear material  refractive index. Highly nonlinear glasses~\cite{Zakery03,Zakery07} and hydrogenated amorphous silicon~\cite{Lacava13,Matres12}, that can be used as a nonlinear medium, have high refractive index $\sqrt{\epsilon_{l,1}}>2$. Therefore, the linear dielectric has to be also a high index material. Consequently, the last goal can not be fulfilled --- it is not possible to access nor measure directly the plasmonic part of the solution if the external layer is filled with a solid. In order to reach this goal, a configuration where the linear refractive index of the external layer is low enough $\sqrt{\epsilon_4} \lesssim 1.3$ needs to be found, so that this external medium can be e.g., water or air. This last problem is solved by the use of four-layer structures, as shown in the next section.

\subsection{Four-layer configuration}
\label{sec:4-lay}

In this section the results obtained with our three models for four-layer configurations are presented. At the beginning, we show and analyze, for the first time, the typical dispersion curve of four-layer configurations. Then the comparison between the results obtained using our three models is performed. The very good agreement between these results confirms the validity of our models. Then, the analysis of the structure parameters is performed and the ranges where low-power plasmon--solitons exist are identified. Later, the advantages of the four-layer structures over three-layer structures are discussed. Finally, the influence of the two geometric parameters of the structure ($d$ and $L$) on the plasmon--solitons properties is presented.

\subsubsection{Dispersion relation}

\begin{figure}[!ht]
\includegraphics[width = 0.238\textwidth,angle=-0,clip=true,trim= 25 40 230 10]{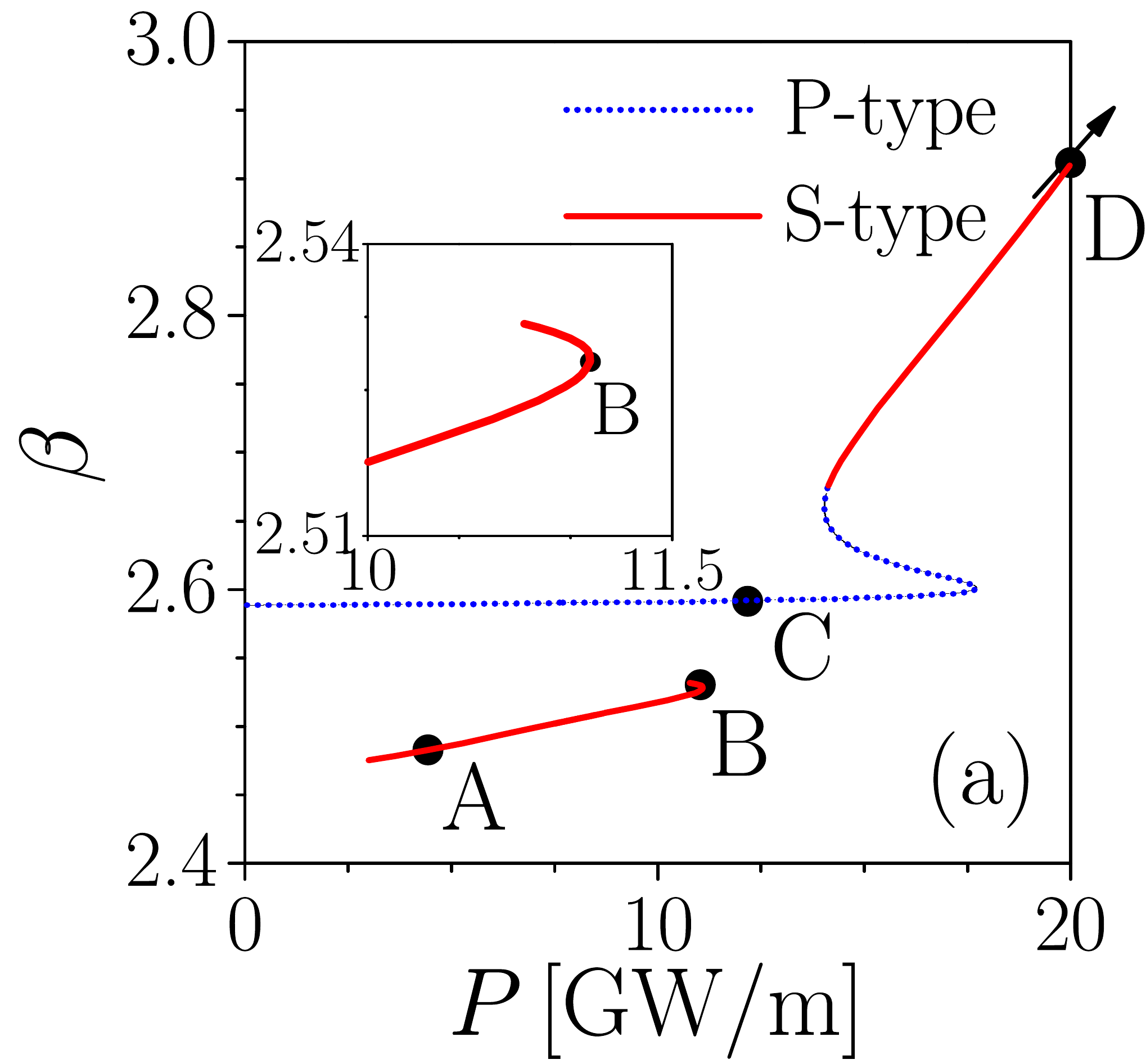}
\includegraphics[width = 0.238\textwidth,angle=-0,clip=true,trim= 4 24 75 3]{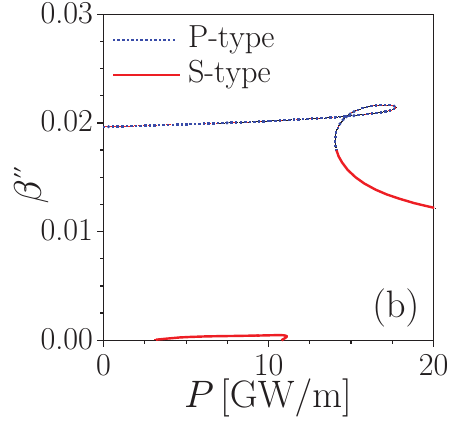}
\caption{Dispersion relation for the real (a) and imaginary parts (b) of the effective index in the four-layer structure with $\epsilon_4 = \epsilon_{l,1}=2.4707$ as a function of power density $P$. Plasmonic type solutions are denoted by a blue dotted line and solitonic type solutions by a red solid line. Point D is located outside the plot boundaries (see the text for explanation). The inset in panel (a) presents the zoom of the lower branch in the vicinity of the point B.}
\label{fig:4-disp}
\end{figure}

The four-layer structure with parameters $\tilde{\epsilon}_{l,1} = 2.4707^2 - 10^{-5}i$, $n_2^{(1)} = 10^{-17}~\rm{m^2/W}$ (chalcogenide glass), $\tilde{\epsilon}_2 = 1.443^2- 10^{-5}i$ (silica), $\tilde{\epsilon}_3 = -96- 10i$ (gold), $\tilde{\epsilon}_4=2.4707^2- 10^{-5}i$ , $L = 15$~nm, $d = 40$~nm, $\lambda = 1.55$~$\mu$m is considered. In Fig.~\ref{fig:4-disp}(a) the dispersion relation $\beta(P)$ for this configuration is presented. There are two separate branches on this plot. The higher branch starts in the linear regime with the plasmonic type solution (P-type, blue dotted curve). With the increase of the power the propagation constant increases. The highest power density of the plasmonic type solution is $P \approx 18$~GW/m. Further increase of the propagation constant is accompanied by the decrease of the power density until $P \approx 14$~GW/m, where another turning point occurs. Slightly above this bend the solution changes its type to solitonic (S-type, red solid curve). The solitonic type solution increases its power with the increase of $\beta$ for the range of $P$ and $\beta$ shown in this plot. The lower branch of the dispersion is purely of the solitonic type. It starts at the level $P\approx 3$~GW/m and the power density increases with the increase of the propagation constant. At $P \approx 11.1$~GW/m, which is the maximum power density for this branch, there is a turning point and $P$ starts to decrease with the increase of $\beta$. The  branch terminates at a power level $P \approx 10.8 $~GW/m. Both ends of the lower branch correspond to $x_0 \rightarrow -\infty$.

In Fig.~\ref{fig:4-disp}(b) the imaginary part of the effective index $\beta''$ is shown. It can be seen that the low-index solitonic type branch is a long range one (it has low losses because the solutions laying on this branch are mainly localized in the nonlinear dielectric). The high index plasmonic type branch and its solitonic type continuation are short range solutions (the high losses of these solutions come from the fact that an important part of the field of these solutions is localized on the lossy metal film).

\begin{figure}[!ht]
\includegraphics[width = 0.238\textwidth,angle=-0,clip=true,trim= 6 20 68 0]{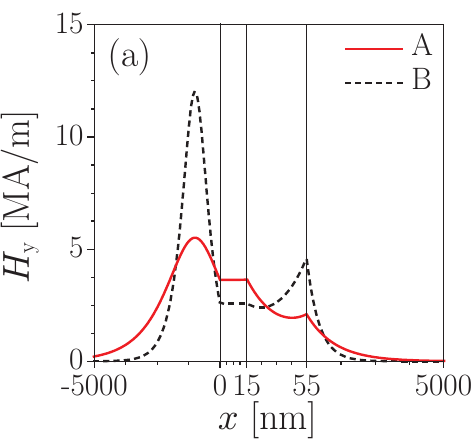}
\includegraphics[width = 0.238\textwidth,angle=-0,clip=true,trim= 5 21 69 2]{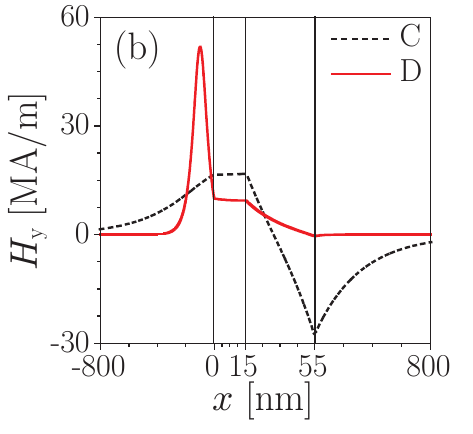}
\caption{Magnetic field profiles corresponding to the solutions marked by points A and B (a) and C and D (b) in Fig.~\ref{fig:4-disp}(a).}
\label{fig:disp-4-fields}
\end{figure}

In Fig.~\ref{fig:disp-4-fields} the characteristic magnetic field profiles corresponding to the points A--D in Fig.~\ref{fig:4-disp}(a) are depicted. In panel (a) the solutions located at the lower branch are presented, both obtained for $x_0 = -1$~$\mu$m. In panel (b) the solutions located at the higher branch are shown. The solitonic type solution (D) was obtained for $x_0 = -0.1$~$\mu$m (the corresponding $\beta = 6.28$ and $P = 45$ GW/cm$^2$) and the plasmonic type solution (C) for $x_0 = 0.1$~$\mu$m.

\subsubsection{Comparison between the results of the three models}
\label{sec:Yin_comp}

\begin{figure}[!ht]
\centering
\includegraphics[width = 0.347\textwidth,angle=-0,clip=true,trim= 25 5 105 5]{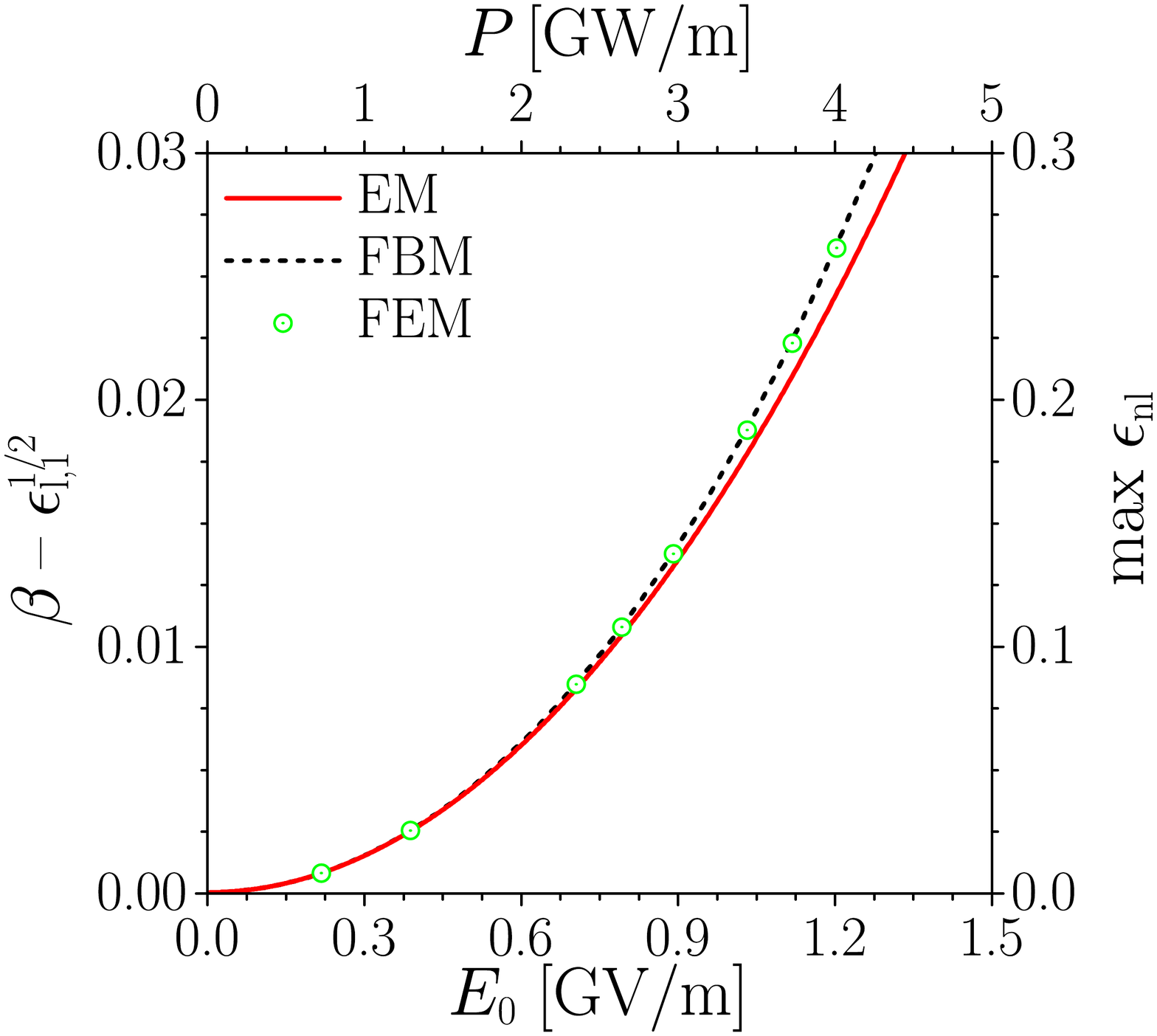}
\caption{Comparison of the nonlinear dispersion relations obtained from: the EM (red solid curve), the FBM (black dashed curve), and the FEM based model (open circles) for a four-layer structure with parameters from Ref~\cite{Walasik12}. The nonlinear variation of the effective index ($\beta - \sqrt{\epsilon_{l,1}}$) is presented in the left vertical axis as a function of the electric field amplitude at the buffer linear dielectric/nonlinear dielectric interface ($x=0$) and of the power density $P$ (on the top axis). On the right vertical axis the maximal nonlinear permittivity change corresponding to the soliton peak is shown.}
\label{fig:4-layer}
\end{figure}

Figure~\ref{fig:4-layer} presents a comparison of the results for the four-layer configuration obtained with the three different models described in section \ref{sec:models}: the FBM, the EM, and the FEM based model. For this comparison the four-layer structure previously presented in Ref.~\cite{Walasik12} is chosen. The parameters of the structure are the same as in the previous paragraph but the external permittivity is now set to $\epsilon_4=1$ (air). Here, only the lowest branch of solitonic type solutions in this structure is presented for relatively low powers. 

First, the results provided by the two semi-analytical models are compared. For the low field amplitudes at the interface between layers 1 and 2 [defined by Eq.~(\ref{eqn:int-amp})] $E_0 \lesssim 0.75$~GV/m, and therefore low maximal nonlinear permittivity change ($\epsilon_{\textrm{nl}} \lesssim  0.1$), both models are in a very good agreement. For higher values of $E_0$ the discrepancy between the FBM and the EM appears. This discrepancy can be explained by looking at the assumptions that were used to built the models. As described in section \ref{sec:field_model} the FBM was formulated by assuming that the nonlinear refractive index changes are small. In this case it is possible to neglect the longitudinal component of the electric field ($E_z$) in the nonlinear contribution to the permittivity, because it is much smaller than the transverse component ($E_x$). For higher nonlinear index modifications both fields contribute with a comparable weight to the nonlinear effects. This is why the results of the FBM differ from those obtained with the EM, that takes both electric field components into account. 
The highest maximal permittivity change shown in Fig.~\ref{fig:4-layer} is of the order of 0.3. Even for such high $\epsilon_{\textrm{nl}}$ the electric field component ratio is $E_x/E_z \approx 10/1$. This justifies the assumption used in the FBM that allowed us to neglect the longitudinal field  in the nonlinear contribution to the permittivity. The maximal relative difference between the results provided by the two models for the effective index variation $\beta - \sqrt{\epsilon_{l,1}}$ is of the order of 10\% for $E_0 \approx 1.4$~GV/m.

The results of the FEM based model shown in Fig.~\ref{fig:4-layer} overlap with the FBM results. This is due to the choice made for the used FEM algorithm which takes into account only the transverse component of the electric field while computing the nonlinear effects. The FEM method solves numerically the nonlinear wave equation [Eq.~(\ref{eqn:wave4})] which is the heart of the FBM. For these reasons it is understandable that this model nicely reproduces the results of the FBM.

\begin{figure}[h!]
  \centering
\includegraphics[width = 0.238\textwidth,angle=-0,clip=true,trim= 25 25 240 15]{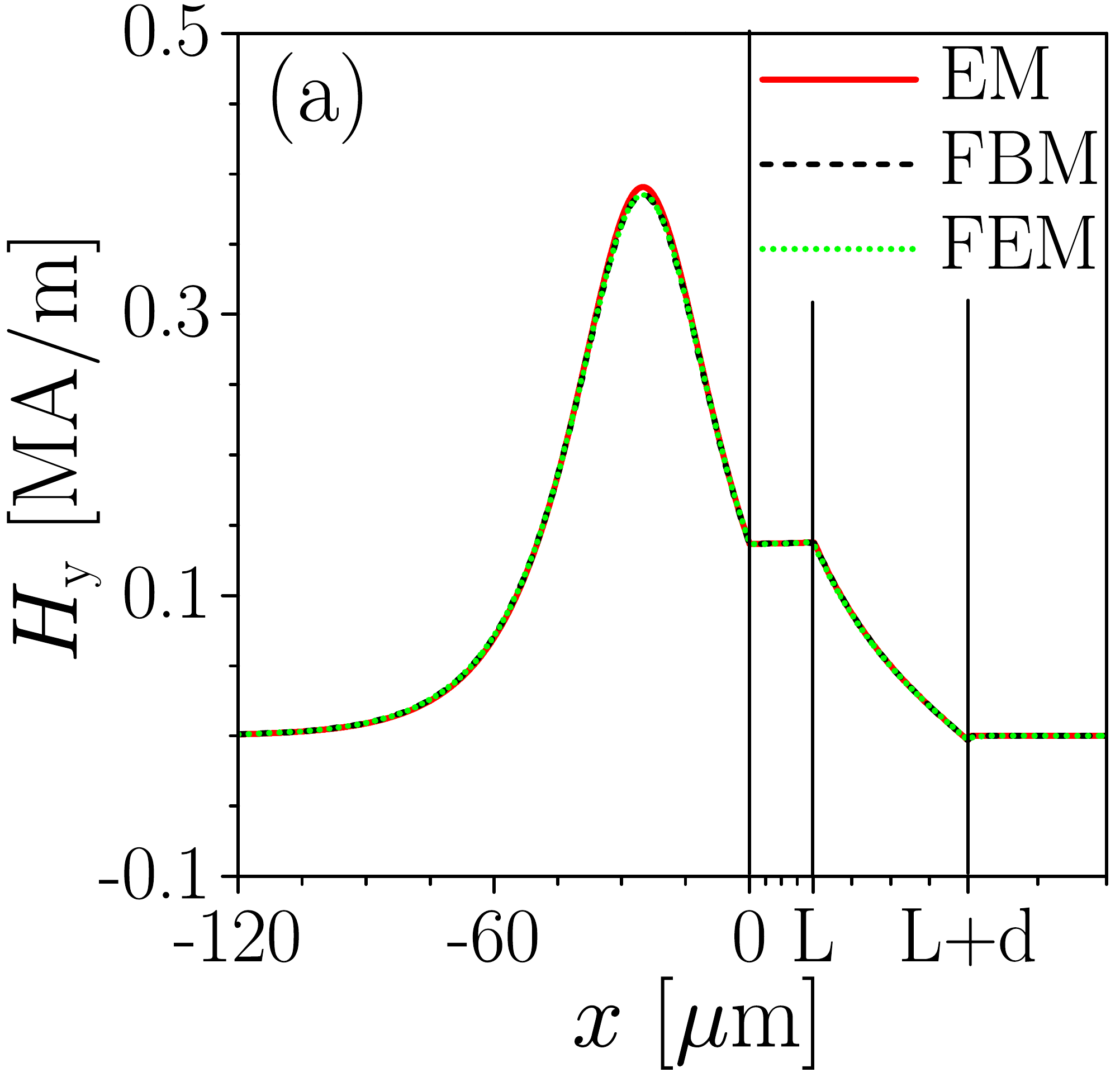}
\includegraphics[width = 0.238\textwidth,angle=-0,clip=true,trim= 25 25 240 15]{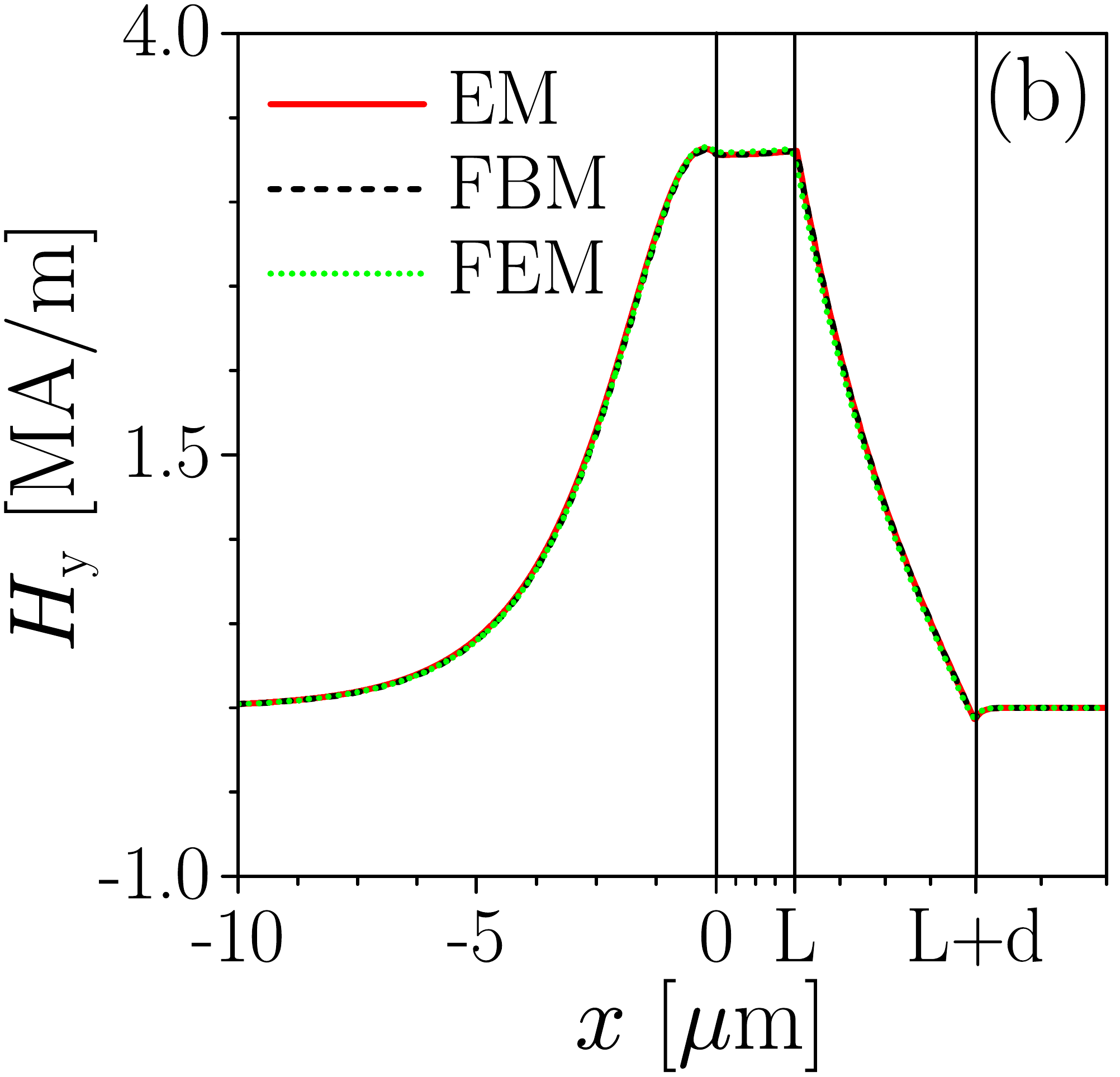}\\
\includegraphics[width = 0.238\textwidth,angle=-0,clip=true,trim= 25 25 240 15]{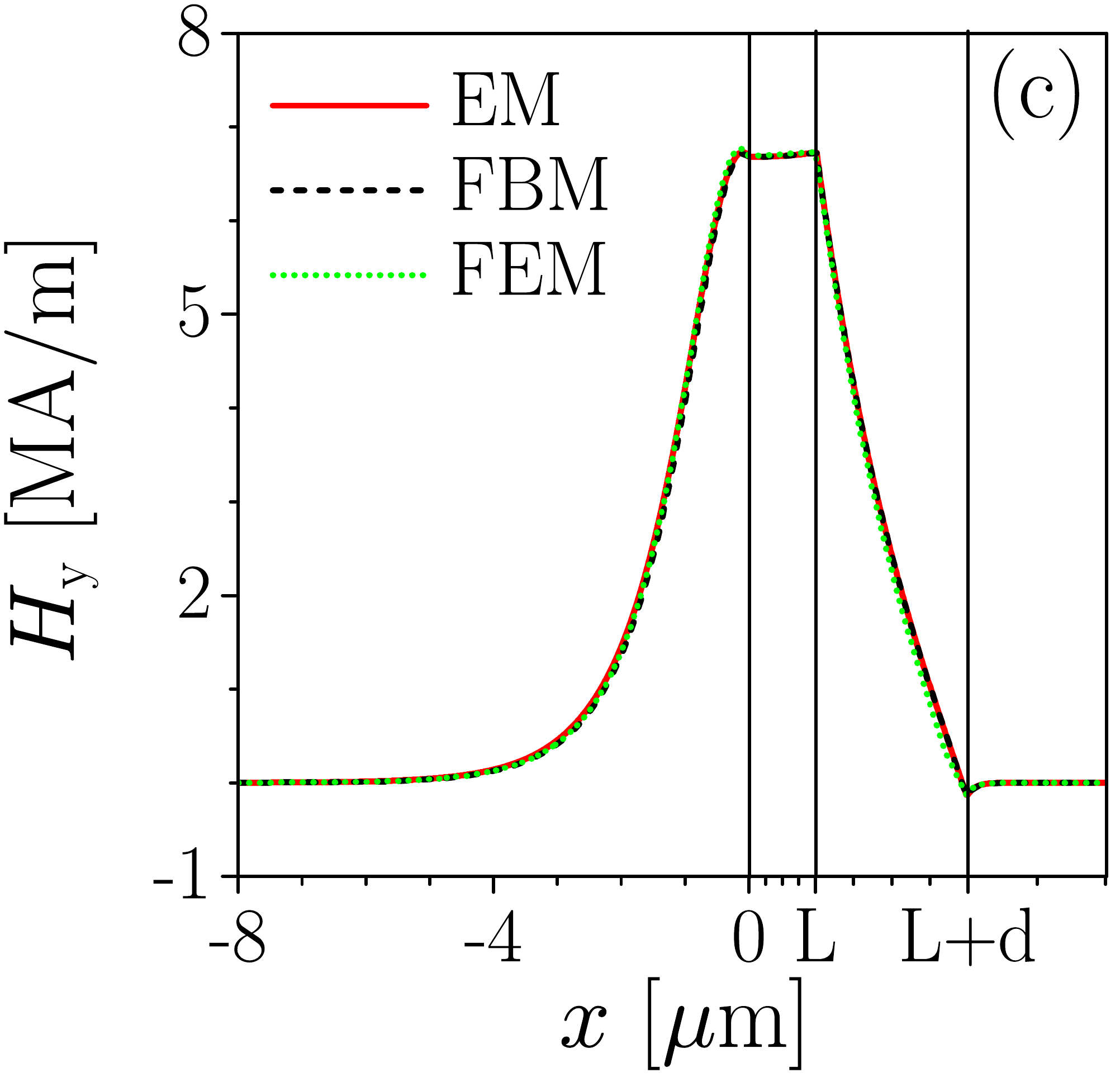}
\caption{Comparison of the magnetic field profiles obtained with the EM (red solid curve), the FBM (black dashed curve), and the FEM based method (green dotted curve) for $E_0$ values: 0.02~GV/m (a), 0.5 GV/m (b) and 1 GV/m (c).}
\label{fig:field_comp}
\end{figure}

In Fig.~\ref{fig:field_comp} the comparison of the field shapes obtained using our three models is presented. Only the $H_y$ field component is shown because all the important observations can be made using this component. The analysis of the electric field components  $E_x$ and $E_z$ does not lead to any new conclusions, and consequently it is omitted. As described in section \ref{sec:yin_disp_field} the field shapes in the nonlinear layer in the EM are not given by an analytical formula but are described by the system of the first order differential equations [Eqs.~(\ref{eqn:yin1}) and (\ref{eqn:yin_fields_system2})]. This system is solved using the $4^{th}$ order Runge--Kutta method~\cite{Press07}. The boundary conditions, allowing to solve this system of equations, take into account the values of the electric field components ($E_{x,0}$ and $E_{z,0}$) at the interface between the nonlinear dielectric and the buffer linear dielectric film (layer 2 in Fig.~\ref{fig:geometry}). These values are found for a given value of $E_0$ using Eqs.~(\ref{eqn:Ez}) and (\ref{eqn:int-amp}). In Fig.~\ref{fig:field_comp}(a) the field profiles for
$E_0 = 0.02$~GV/m are presented. In panel (b) $E_0 = 0.5$~GV/m and in panel (c) $E_0 = 1$~GV/m. In all the cases, the fields obtained with the FBM and the FEM based method are in a very good agreement. The fields obtained using the EM also overlap very well with the previous ones despite the small discrepancies of the corresponding propagation constants.

\subsubsection{Toward low-power solutions}

\label{sec:res-4-lay-general}

In order to find low-power solitonic type solutions for the configurations with a high index contrast between the nonlinear dielectric and the linear external dielectric, the properties of four-layer configurations are investigated. 

In this section, the parameters used to obtain all the color maps (two-parameter scans performed with the FBM) are the same as in Ref.~\cite{Walasik12} and in section \ref{sec:Yin_comp} except if explicitly stated or if the parameters are on the axes of the plot.
We have chosen $x_0=-15.5$~$\mu$m value for all the illustrations.
In all the plots only the effective indices from the range $\sqrt{\epsilon_{l,1}} < \beta <4 \sqrt{\epsilon_{l,1}}$ are shown (like in section~\ref{sec:3-lay-low-search}). 

\begin{figure}[!ht]
  \centering
\includegraphics[width = 0.238\textwidth,angle=-0,clip=true,trim= 7 24 60 3]{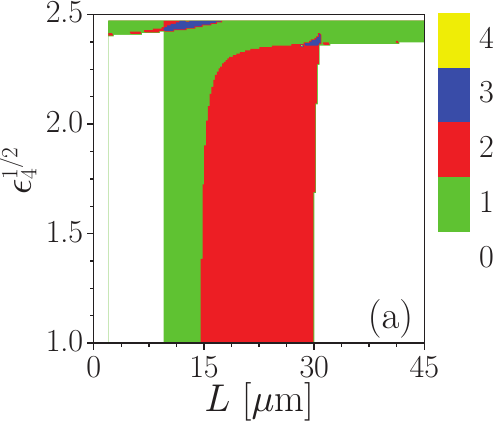}
\includegraphics[width = 0.238\textwidth,angle=-0,clip=true,trim= 9 24 58 3]{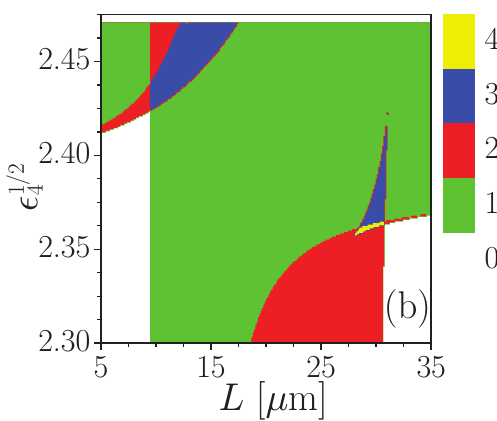}
\caption{(a) Number of solitonic type solutions in a four-layer structure as a function of the buffer layer thickness $L$ and of the external layer refractive index $\sqrt{\epsilon_4}$ and (b) the zoom on the most complex part of the plot.}
\label{4l-nb_L_eps4}
\end{figure}

First, the evolution of the number of solitonic type solutions as a function of the  linear buffer layer thickness $L$ and of the external layer refractive index $\sqrt{\epsilon_4}$ is analyzed. It is seen from Fig.~\ref{4l-nb_L_eps4} that for low buffer layer thickness $0<L\lesssim9$~nm the four-layer structure presents similar behavior as the three-layer structure (see Figs.~\ref{3_lay_nb_power} and \ref{3_lay_nb_met}). There is one solitonic type solution for the quasi-symmetric case $n_{\textrm{cut-off}} \approx 2.4 < \sqrt{\epsilon_4} < \sqrt{\epsilon_{l,1}}$ and no solitonic type solutions for higher index contrasts between the external layer and the nonlinear dielectric. These two cases are separated by a narrow region with two solutions, that becomes broader with the increase of the buffer thickness [see Fig.~\ref{4l-nb_L_eps4}(b)]. For buffer thickness between $9$~nm and $30$~nm there is up to three solitonic type solutions possible for low index contrast regime $\sqrt{\epsilon_4} > n_{\textrm{cut-off}}$ an even up to four solutions [yellow region in Fig.~\ref{4l-nb_L_eps4}(b)] in a small region for a moderate index contrast configuration. For the buffer thickness above $30$~nm only a single solitonic type solution exists in low and moderate index contrast regimes.

In the region with three or four solitonic type solutions occurring for the same $x_0$ value, two of the corresponding field shapes are analogous to those presented in Fig.~\ref{fig:disp-4-fields}(a). The other solutions have even higher effective indices $\beta$ and therefore even narrower solitonic parts and higher peak powers than the two previously mentioned solutions.
\begin{figure}[!ht]
  \centering
\includegraphics[width = 0.238\textwidth,angle=-0,clip=true,trim= 15 40 170 10]{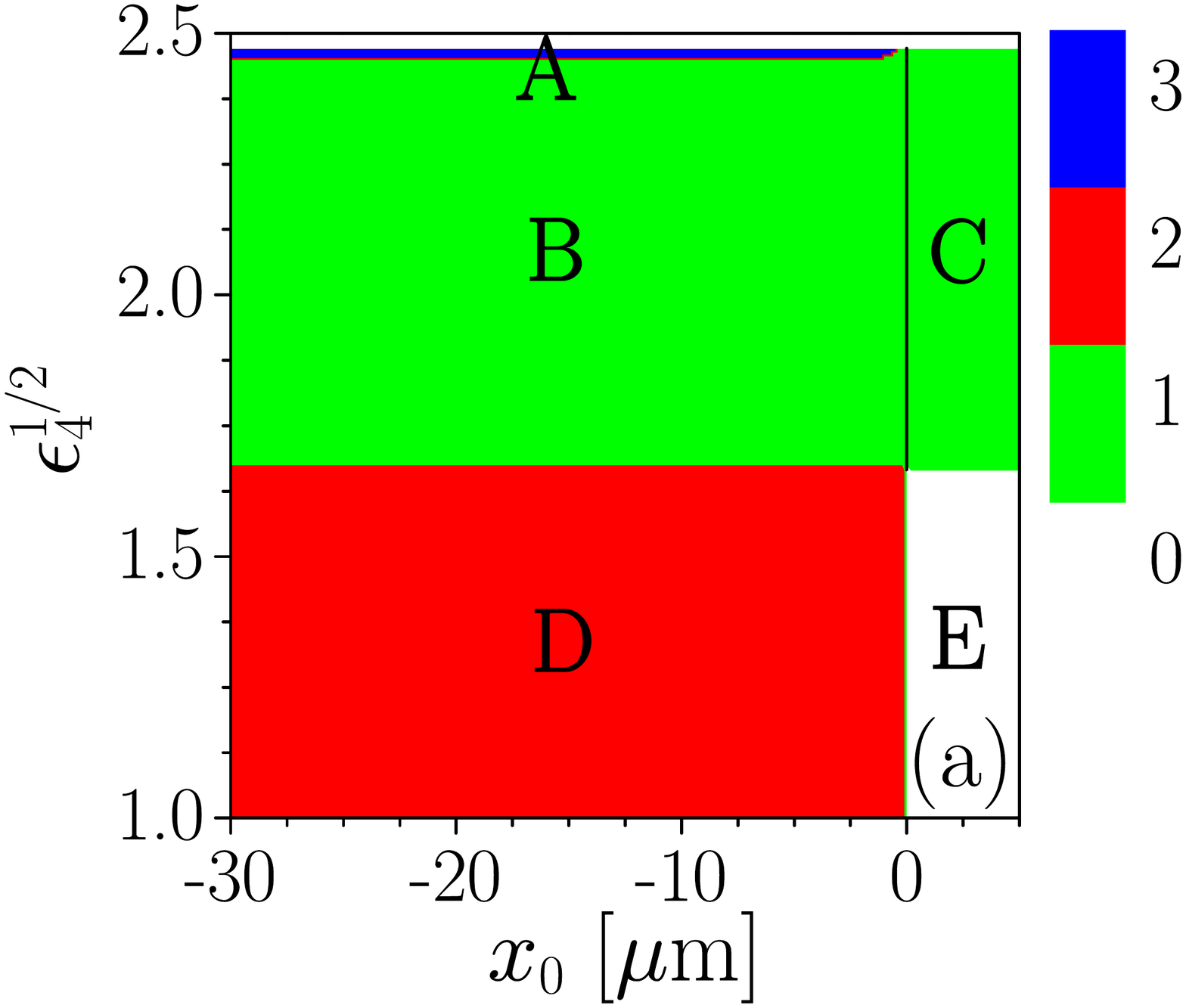}
\includegraphics[width = 0.238\textwidth,angle=-0,clip=true,trim= 25 40 160 10]{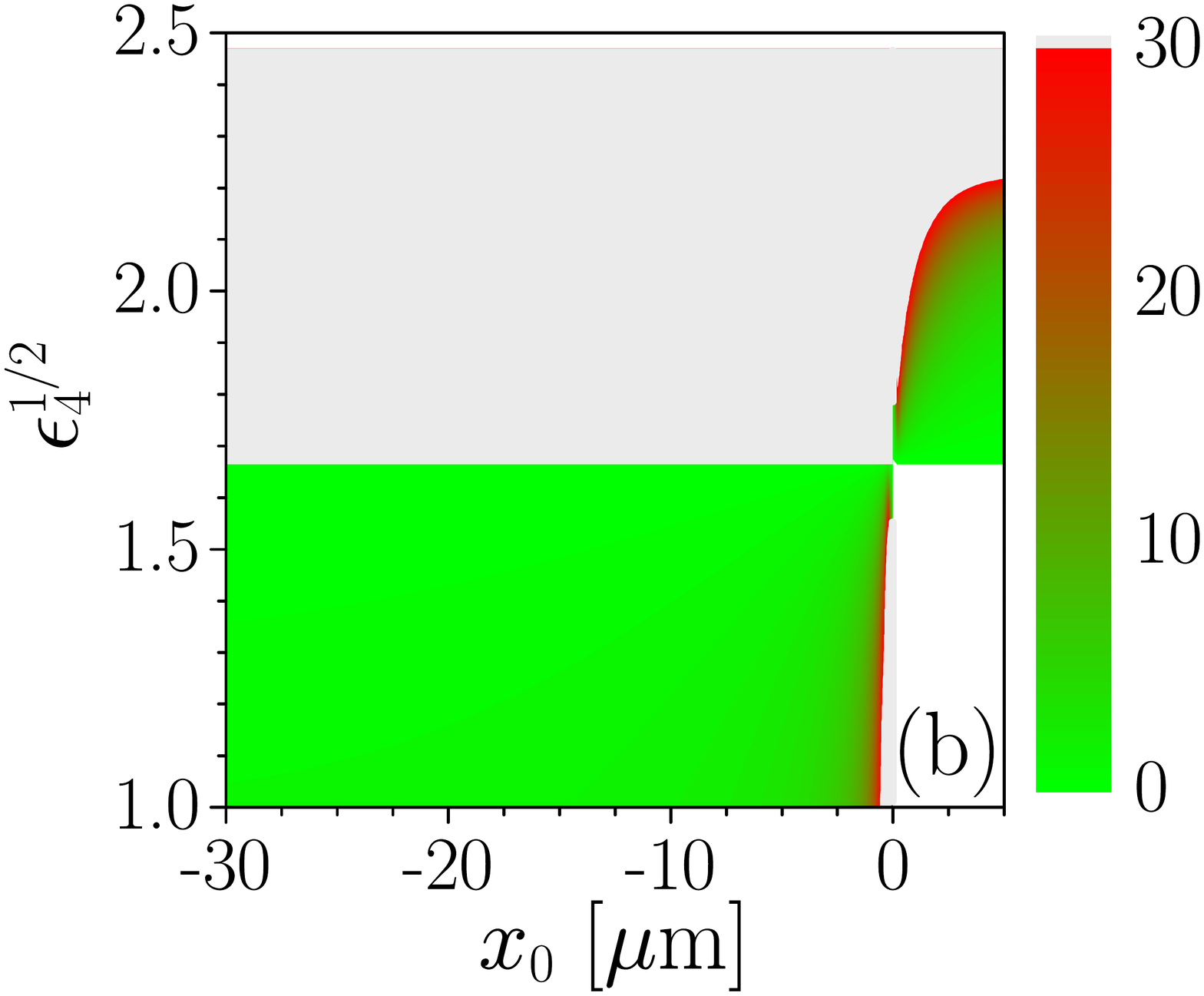}
\caption{(a) Number of solutions in a four-layer structure as a function of the external layer refractive index $\sqrt{\epsilon_4}$ and of the parameter $x_0$. (b) Peak power [GW/cm$^2$]  for the low-power solutions.  The existence of solutions with higher peak power is marked with the gray color}.
\label{4l-nb_x0_eps4}
\end{figure}

In Fig.~\ref{4l-nb_x0_eps4}(a) we show the total number of solutions as a function of the external layer refractive index $\sqrt{\epsilon_4}$ and of the $x_0$ parameter [in analogy to Fig.~\ref{3_lay_nb_power}(a) for three-layer structures]. In this case we see that in a quasi-symmetric structure ($\epsilon_4 \approx \epsilon_{l,1}$) there are three (region A) or two (for $x_0$ values close to zero) solitonic type solutions and one plasmonic type solution (top of the region C). For the region with a moderate index contrast ($1.7\lesssim \sqrt{\epsilon_4} \lesssim2.4$) there is one solitonic type solution (region B) and one plasmonic type solution (region C). Finally for high index contrast  ($\sqrt{\epsilon_4} \lesssim1.7$) there exist two solitonic type solutions (region D) and no plasmonic type solution (region E). The value of $\sqrt{\epsilon_4} \approx1.7$ is a cut-off limit both in the case of solitonic and plasmonic type solutions. Increasing $\sqrt{\epsilon_4}$ for positive $x_0$ values causes the appearance of a plasmonic type solution. On the other hand, for negative values of $x_0$ this causes a reduction of the number of solitonic type solutions from two to one.

Figure~\ref{4l-nb_x0_eps4}(b) shows the peak power of the solutions in four-layer configurations. Similarly to the three-layer case [shown in Fig.~\ref{3_lay_nb_power} (b)] the lowest peak intensities occur below the cut-off index for solitonic type solutions and above this value for plasmonic type solutions. However, in this case, for plasmon--solitons the region of low-power solutions extends to much lower external layer refractive indices than in the case of a three-layer configuration. This means that in a four-layer configuration we are not only able to find plasmon--solitons for high index contrast configurations but also that these solutions have low peak intensities. 

It must be pointed out that the maps presented in Fig.~\ref{4l-nb_x0_eps4} have been obtained for a value of $L = 15$~nm which corresponds to a cut in a relatively simple region of the map provided in Fig.~\ref{4l-nb_L_eps4}. More complicated maps can be obtained  for specific $L$ values (e.g. $L=28$~nm --- data not shown) but the obtained nonlinear solutions still belong to the classification provided in section~\ref{sec:mod_class}.

\begin{figure}[!ht]
  \centering
\includegraphics[width = 0.238\textwidth,angle=-0,clip=true,trim= 15 40 170 10]{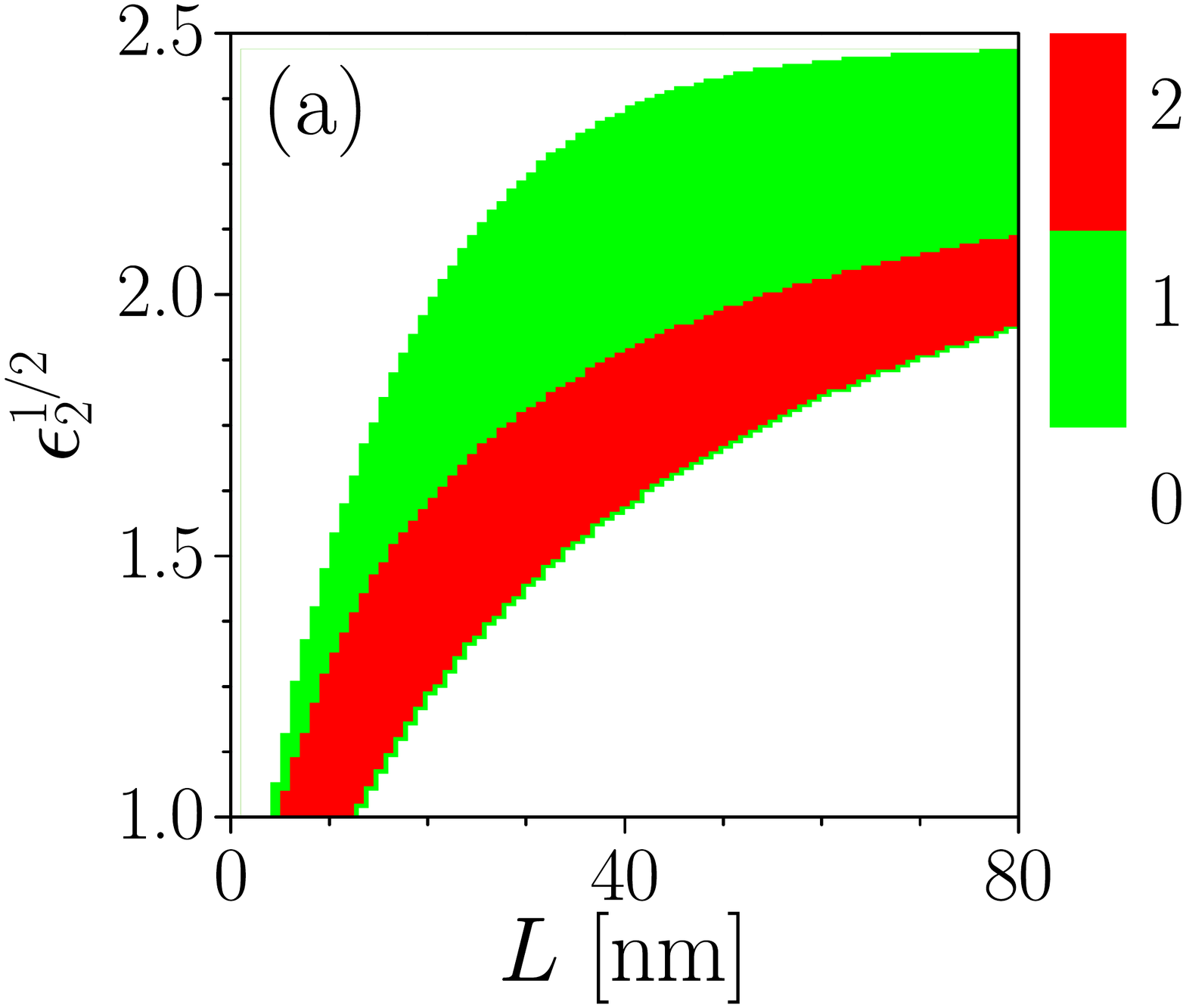}
\includegraphics[width = 0.238\textwidth,angle=-0,clip=true,trim= 25 40 160 10]{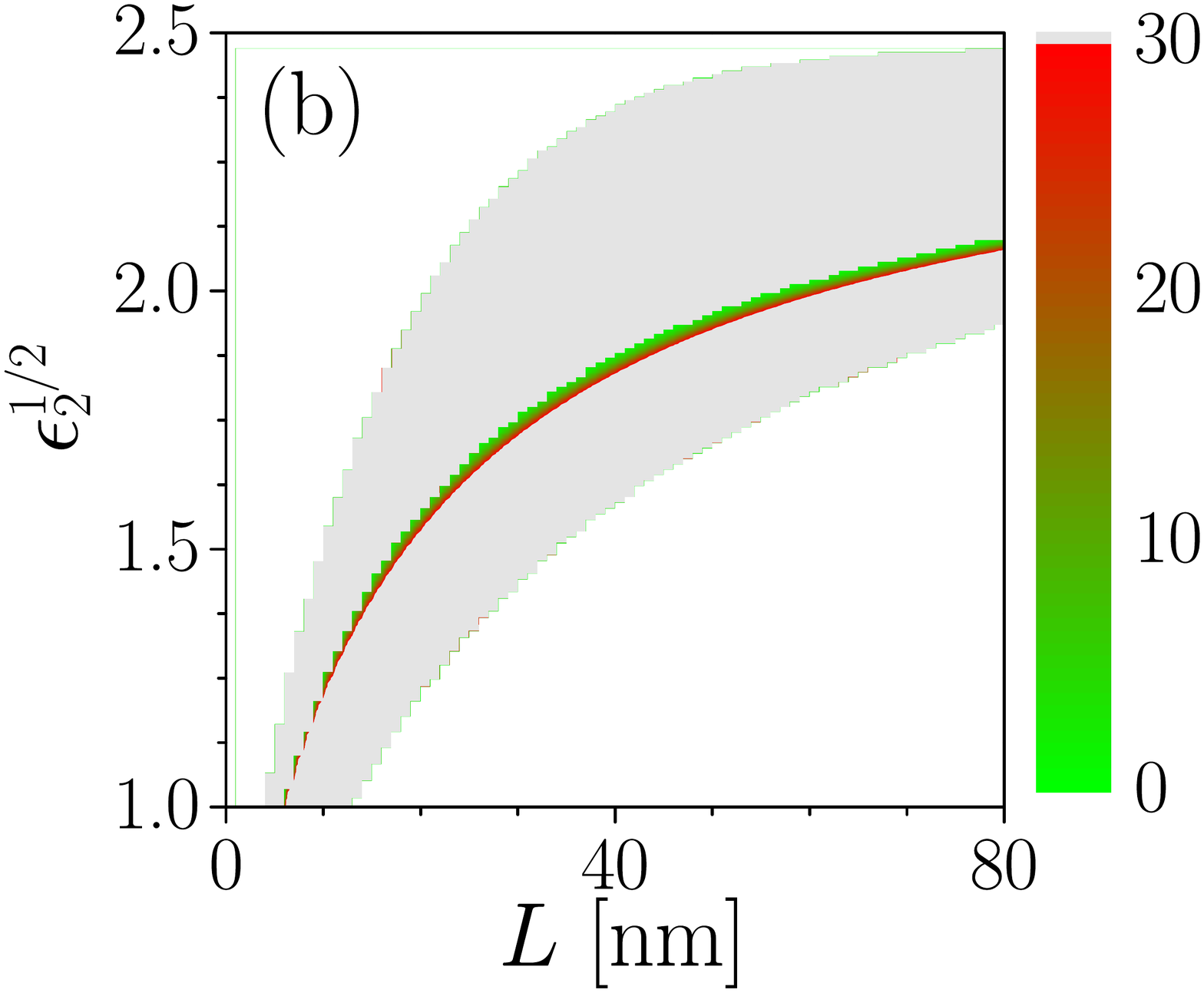}
\caption{(a) Number of solitonic type solutions as a function of the buffer layer thickness $L$ and of the refractive index of this layer $\sqrt{\epsilon_2}$. (b) Peak power [GW/cm$^2$]  for the low-power solutions.} 
\label{4l-nb_L_eps2}
\end{figure}

Figure~\ref{4l-nb_L_eps2}(a) shows the number of solitonic type solutions as a function of the buffer layer thickness $L$ and of the refractive index of this layer $\sqrt{\epsilon_2}$. It can be seen that for low buffer layer refractive index ($\sqrt{\epsilon_2}=1$) the range of thickness where one or two solutions exist is quite narrow (5--15~nm). Increasing the buffer layer refractive index, the range of the buffer thickness where the solutions exist expands (it becomes approximately 45--80~nm for $\sqrt{\epsilon_2} = 1.75$).

Figure~\ref{4l-nb_L_eps2}(b) shows the plasmon--soliton peak power in the same coordinates as those used in panel (a). The region where plasmon--solitons have low peak intensities is very narrow and is located close to the line separating regions with one and two solutions. Increasing the buffer layer refractive index allows an increase of the buffer layer thickness required to obtain solutions with low peak power, which is interesting from a technological point of view (e.g.~it is challenging to fabricate uniform, high quality thin films on top of chalcogenide glasses~\cite{Nazabal08}).

\begin{figure}[!ht]
\centering
\includegraphics[width = 0.238\textwidth,angle=-0,clip=true,trim= 9 24 52 3]{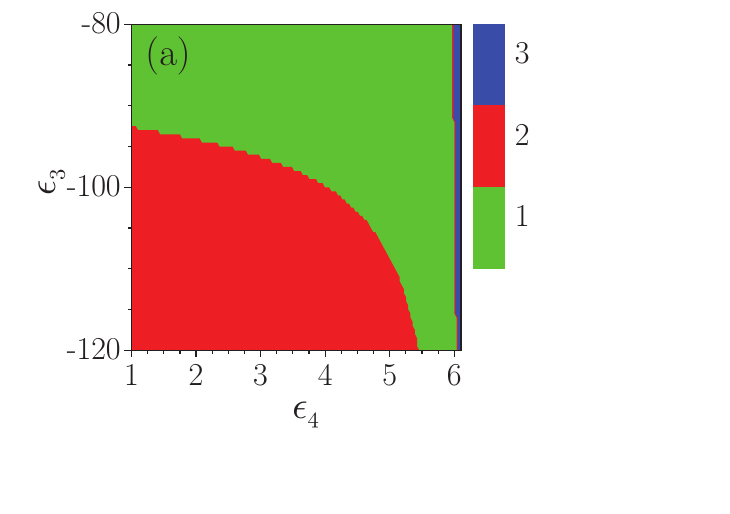}
\includegraphics[width = 0.238\textwidth,angle=-0,clip=true,trim= 5 24 56 3]{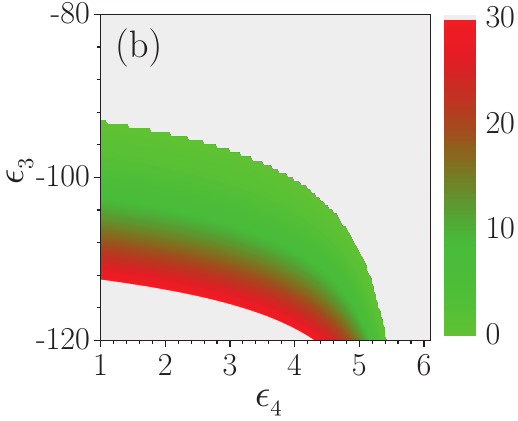}
\caption{(a) Number of solitonic type solutions as a function of the metal layer permittivity $\epsilon_3$ and of the external medium permittivity $\epsilon_4$. (b) Peak power [GW/cm$^2$]  for the low-power solutions.}
\label{4l-nb_eps_2_eps_4}
\end{figure}

Figure~\ref{4l-nb_eps_2_eps_4}(a) presents the number of solitonic type solutions as a function of the metal layer permittivity $\epsilon_3$ and of the external medium permittivity $\epsilon_4$ (it can be compared with Fig.~\ref{3_lay_nb_met}(a) presenting the analogous dependency for a three-layer structure). The main advantage of the four-layer structure compared to the three-layer one is that, even for very low permittivity of the external medium (like 1 for air or $1.3^2$ for water at $\lambda = 1.55$~$\mu$m) resulting in high index contrast, the solitonic type solutions exist. There is two of them for low metal permittivity values and one for higher metal permittivity values. In four-layer structures where $\epsilon_4 \approx \epsilon_{l,1}$ even three solitonic type solutions exist for the same $x_0$ parameter [the blue region in Fig.~\ref{4l-nb_eps_2_eps_4}(a)].

Figure~\ref{4l-nb_eps_2_eps_4}(b) shows the peak power of the solitonic type solutions in the same coordinates as those used in panel (a). Comparing this figure with the corresponding one for a three-layer structure in Fig.~\ref{3_lay_nb_met} (b) it can be seen that in the case of four-layer configuration low-power solutions exist for wider ranges of both $\epsilon_3$ and $\epsilon_4$ which broadens the choice of possible parameter combinations. This property may facilitate the fabrication of the structure.

\subsubsection{Optimization of the four-layer structure}

\begin{figure}[!ht]
  \centering
\includegraphics[width = 0.238\textwidth,angle=-0,clip=true,trim= 7 23 60 2]{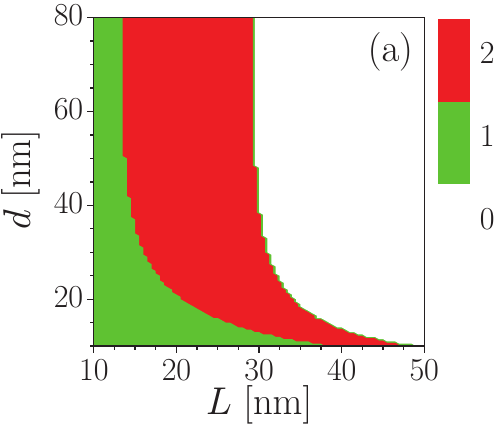}
\includegraphics[width = 0.238\textwidth,angle=-0,clip=true,trim= 8 23 59 2]{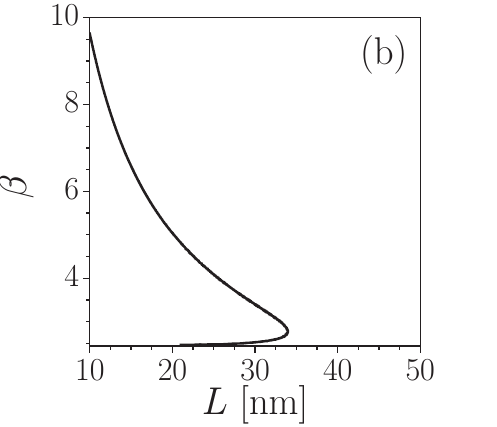}
\caption{(a) Number of solitonic type solutions as a function of the metal film thickness $d$ and of the buffer layer thickness $L$. (b) The effective index $\beta$ as a function of the buffer layer thickness $L$ for a fixed metal thickness $d=20$~nm.}
\label{4l-nb_L_d-eps-converted-to.pdf}
\end{figure}

In this section, a more detailed investigation of the influence of the two geometrical parameters of the four-layer structure (the metal layer thickness $d$ and the buffer layer thickness $L$) is shown. Figure \ref{4l-nb_L_d-eps-converted-to.pdf}(a) shows the number of solitonic type solutions as a function of these two parameters. For low values of the thickness of both layers only one solution is obtained. For higher values of dielectric buffer thickness there exist a region for which two solutions appear. For even higher values of $L$ both solutions disappear. The evolution of the solutions can be followed by looking at Fig.~\ref{4l-nb_L_d-eps-converted-to.pdf}(b) which corresponds to a cut of Fig.~\ref{4l-nb_L_d-eps-converted-to.pdf}(a) at $d=20$~nm. For low values of $L$ only a high effective index solution exists. $L\approx21$~nm is a cut-off buffer thickness for a second solitonic type solution. At this thickness a low effective index solution appears. As the buffer layer thickness increases, these two solutions become closer to each other to finally merge into one solution for a particular value of $L \approx 34$~nm. Above this value no solitonic type solution exists.

\begin{figure}[!t]
  \centering
\includegraphics[width = 0.238\textwidth,angle=-0,clip=true,trim= 7 21 50 3]{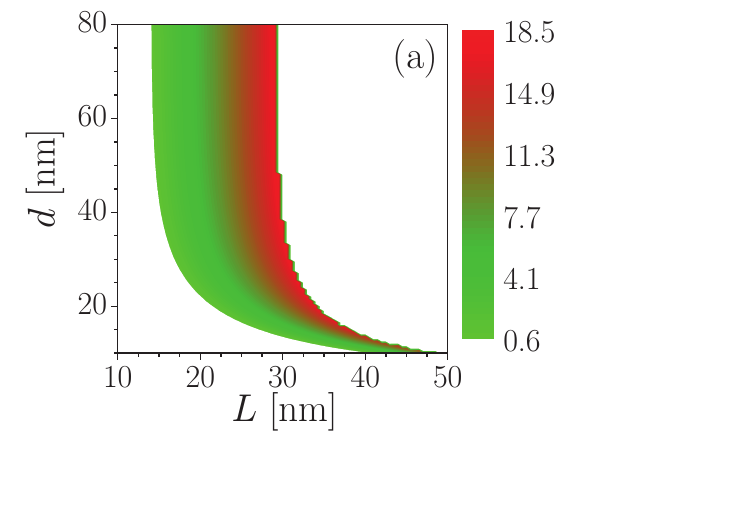}
\includegraphics[width = 0.238\textwidth,angle=-0,clip=true,trim= 7 21 50 3]{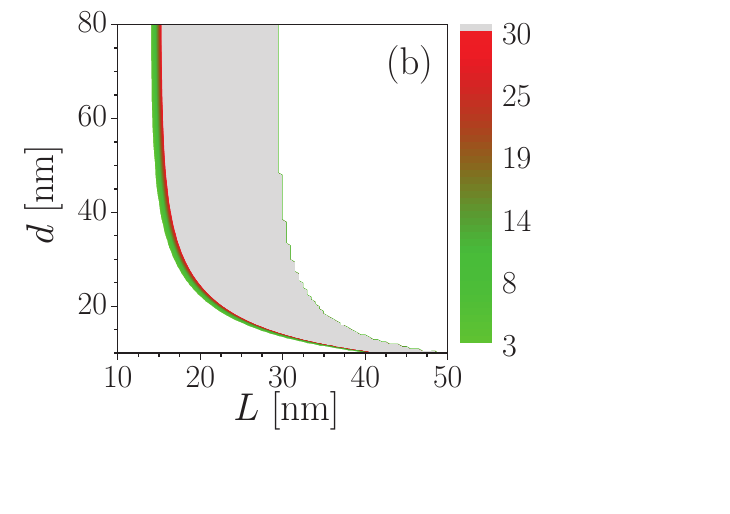}
\caption{(a) Total power density $[\rm{GW/m}]$ of the low $\beta$ solitonic type solution. (b) Peak power [GW/cm$^2$] for the low-power solutions.}
\label{4l-pow_dens_L_d}
\end{figure}

In Fig.~\ref{4l-pow_dens_L_d}(a) the total power density for the solitonic type solution with the lower $\beta$ is shown in the same coordinates as those used in Fig.~\ref{4l-nb_L_d-eps-converted-to.pdf}(a). The solutions with the lowest power density are located close to the cut-off buffer thickness [left region of Fig.~\ref{4l-pow_dens_L_d}(a)].  In Fig.~\ref{4l-pow_dens_L_d}(b) the peak power for the low-power solutions is shown. Plasmon--solitons with the lowest peak intensities are located in a narrow region where the total power density is the lowest (i.e. close to the cut-off buffer thickness $L$ for the low-power solution). This shows that in order to obtain solutions with the peak power levels that are attainable by modern high-power commercial lasers the couple $L$ and $d$ has to be precisely chosen. Even small deviation of the buffer film thickness (e.g.~2~nm) may lead to the change of the peak power of the supported solution by one order of magnitude (e.g.~from $3$ to $30$~GW/cm$^2$).

\begin{figure}[!ht]
  \centering
\includegraphics[width = 0.238\textwidth,angle=-0,clip=true,trim= 25 40 160 10]{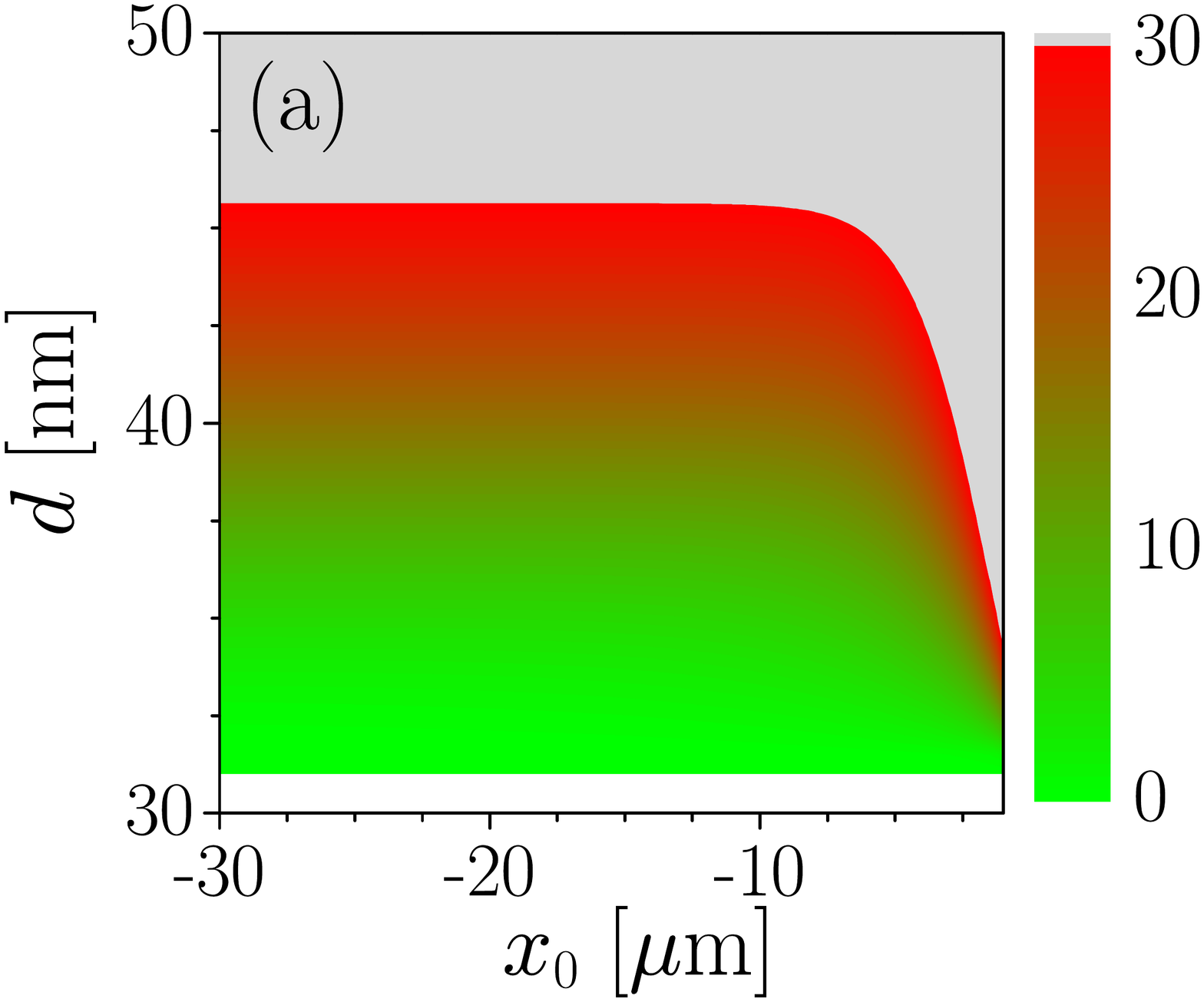}
\includegraphics[width = 0.238\textwidth,angle=-0,clip=true,trim= 25 40 160 10]{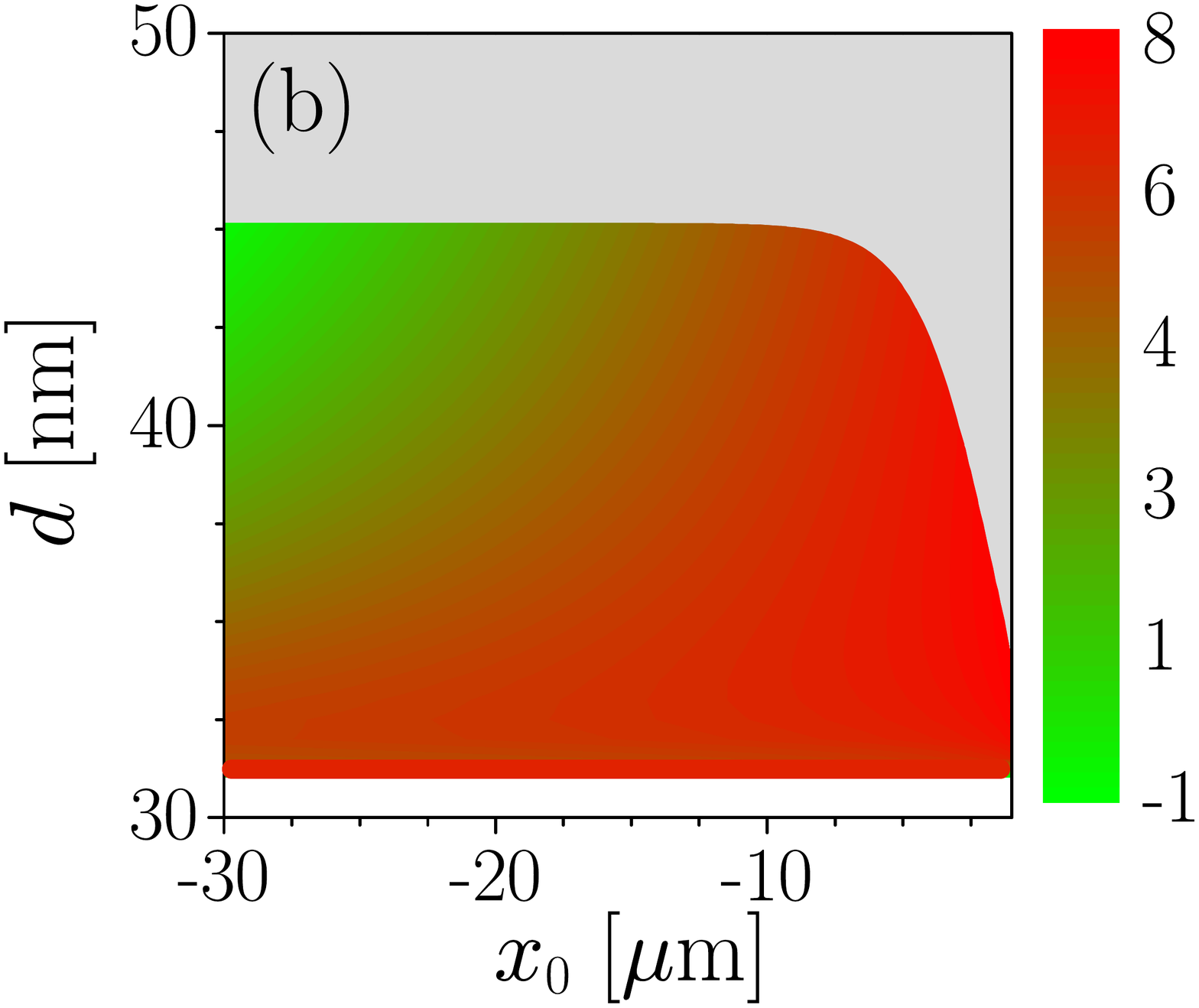}
\caption{(a) Peak power of the solitonic part [GW/cm$^2$] and (b) decimal logarithm of the peak power of the plasmonic part $I_{\textrm{plas}} \equiv I(x=L+d)$: $\log_{10}[I_{\textrm{plas}}$/(Wcm$^{-2})$] for the low-power solitonic type solutions as a function of the metal thickness $d$ and of the parameter $x_0$ for the buffer thickness fixed at $L=16$~nm. In color solutions with solitonic part peak power below 30 GW/cm$^2$ are plotted.}
\label{4l-peak_I_x0_d}
\end{figure}

Figure~\ref{4l-peak_I_x0_d}(a) shows the dependency of the peak power of the solitonic part of the solution as a function of the metal thickness $d$ and $x_0$ parameter for a fixed buffer thickness $L=16$~nm. With the increase of the metal thickness or with the increase of the $x_0$ parameter (decrease in absolute value) the peak power of the solitonic part increases. Besides the peak power of the solitonic part, that should be kept low, there is another important parameter that should be taken into account. It is interesting to have a strong plasmonic field at the interface between the metal film and the external medium in order to facilitate its recording or to make use of it. Figure~\ref{4l-peak_I_x0_d}(b) shows the decimal logarithm of the maximum peak power of the plasmonic part as a function of the metal thickness $d$ and $x_0$ parameter. The lowest values of plasmonic part peak power are obtained for thick metal film and solitonic peak located far from the metal interface. On one hand for large metal thickness values, bringing the solitonic part closer to the metal interface results in a drastic increase (few orders of magnitude) of the peak power in the external layer. On the other hand, for thin metal films the peak power in the external layer is relatively high and the changes with the $x_0$ parameter are much slower.

\section{Conclusions}

We have presented three complementary models based on Maxwell's
equations to study the properties of the stationary TM solutions in planar
nonlinear structures containing a metal film, a semi-infinite nonlinear medium of the Kerr type, and possibly a semi-infinite linear dielectric and a linear dielectric film. 
Two of these models are semi-analytical and allow for fast opto-geometric parameter space scanning in order to find all the possible stationary nonlinear
solutions. They are the extended versions of known models to more complex structures.
The field based model additionally improves the approximation used previously to deal with the nonlinearity.
The other semi-analytical model provides the exact treatment of the nonlinear term involving both the transverse and longitudinal components of the electric field and can be used even for high nonlinear index modifications.  

A more numerical approach  based on a finite element method confirmed the results obtained from the two semi-analytical approaches. Our results also agree with previous
results already published for particular cases of simpler structures.    

A systematic study
of two-, three-, and four-layer configurations was performed that led to
several conclusions. 
Firstly, three main types of nonlinear solutions are found in these structures: symmetric type nonlinear plasmons,
 antisymmetric type nonlinear plasmons, and plasmon--solitons that exhibit a maximum, local or not, of the transverse electromagnetic field components in the nonlinear layer.  
Secondly, the simplest structures supporting plasmon--solitons
with pronounced soliton peak  are found to be composed of three layers (semi-infinite nonlinear Kerr medium/metal film/semi-infinite dielectric). Low-power
solutions are found in three-layer structures only in the case of small refractive index
contrast between the nonlinear and the external dielectrics. 
Thirdly, we show that to overcome this limitation, four-layer
structures  (semi-infinite nonlinear Kerr medium/dielectric film/metal film/semi-infinite dielectric) must be considered.  These structures support low peak power plasmon--solitons even for high index contrast between the two outer dielectrics. 
 Comparison between three- and four-layer structures revealed that the parameter regions for which low-power solutions are obtained are much broader in the latter case. This shows that the constraints for the parameters of the structure that supports low-power solutions are relaxed in the case of four-layer configurations which is desirable from the technological point of view.
Since  physically realistic parameters were used, our results indicate that
the experimental observation of plasmon--solitons should be possible.

\begin{acknowledgments}
This work was supported by the European Commission through the Erasmus Mundus Joint Doctorate Programme  Europhotonics (Grant No. 159224-1-2009-1-FR-ERA MUNDUS-EMJD).
\end{acknowledgments}

\end{document}